\documentstyle[preprint,epsfig,aps]{revtex}
\begin{document}
\tighten

\def\beq{\begin{equation}}
\def\eeq{\end{equation}}
\def\beqy{\begin{eqnarray}}
\def\eeqy{\end{eqnarray}}
\def\pr#1#2#3{ {\sl Phys. Rev.\/} {\bf#1}, #3 (#2)}
\def\prl#1#2#3{ {\sl Phys. Rev. Lett.\/} {\bf#1}, #3 (#2)}
\def\np#1#2#3{ {\sl Nucl. Phys.\/} {\bf#1}, #3 (#2)}
\def\cmp#1#2#3{ {\sl Comm. Math. Phys.\/} {\bf#1}, #3 (#2)}
\def\pl#1#2#3{ {\sl Phys. Lett.\/} {\bf#1}, #3 (#2)}
\def\apj#1#2#3{ {\sl Ap. J.\/} {\bf#1}, #3 (#2)}
\def\aop#1#2#3{ {\sl Ann. Phy.\/} {\bf#1}, #3 (#2)}
\def\nc#1#2#3{ {\sl Nuovo Cimento }{\bf#1}, #3 (#2)}
\def\cjp#1#2#3{ {\sl Can. J. Phys. }{\bf#1}, #3 (#2)}
\def\zp#1#2#3{ {\sl Z. Phys. }{\bf#1}, #3 (#2)}
\def\pbi{\vec{p}_i}
\def\pbj{\vec{p}_j}
\def\yc{{\cal Y}}
\def\tp0{^3P_0}
\def\nii{\noindent}
\def\pbb{{\bf p}}
\def\Pbb{{\bf P}}
\def\qbb{{\bf q}}
\def\kbb{{\bf k}}
\def\Kbb{{\bf K}}
\def\yc{{\cal Y}}
\def\he{\hat \ell}
\def\hs{\hat S}
\def\hl{\hat L}
\def\hj{\hat J}
\def\rmb#1{{\bf #1}}
\def\lpmb#1{\mbox{\boldmath $#1$}}
\def\half{{\textstyle{1\over2}}}
\def\thalf{{\textstyle{3\over2}}}
\def\fhalf{{\textstyle{5\over2}}}
\def\shalf{{\textstyle{7\over2}}}
\def\nhalf{{\textstyle{9\over2}}}
\def\lhalf{{\textstyle{11\over2}}}
\def\tthalf{{\textstyle{13\over2}}}
\def\fthalf{{\textstyle{15\over2}}}
\def\nn{\nonumber}
\def\>{\rangle}
\def\<{\langle}

\title{Strange Decays of Nonstrange Baryons}
\author{Simon Capstick}
\address{Supercomputer Computations Research Institute
  and Department of Physics,\\
  Florida State University,
  Tallahassee, FL 32306}
\author{W. Roberts}
\address{Department of Physics, Old Dominion University\\
Norfolk, VA 23529 USA \\ and \\
Thomas Jefferson National Accelerator Facility\\
12000 Jefferson Avenue, Newport News, VA 23606, USA}
\date{\today}
\maketitle

\begin{flushright}
FSU-SCRI-98-42

JLAB-THY-98-15
\end{flushright}

\begin{abstract}
The strong decays of excited nonstrange baryons into the final states
$\Lambda K$, $\Sigma K$, and for the first time into $\Lambda(1405)
K$, $\Lambda(1520) K$, $\Sigma(1385) K$, $\Lambda K^*$, and $\Sigma
K^*$, are examined in a relativized quark pair creation model. The
wave functions and parameters of the model are fixed by previous
calculations of $N \pi$ and $N \pi\pi$, {\it etc.}, decays. Our
results show that it should be possible to discover several new
negative parity excited baryons and confirm the discovery of several
others by analyzing these final states in kaon production
experiments. We also establish clear predictions for the relative
strengths of certain states to decay to $\Lambda(1405)K$ and
$\Lambda(1520)K$, which can be tested to determine if a three-quark
model of the $\Lambda(1405)K$ is valid. Our results compare favorably
with the results of partial wave analyses of the limited existing data
for the $\Lambda K$ and $\Sigma K$ channels. We do not find large
$\Sigma K$ decay amplitudes for a substantial group of predicted and
weakly established negative-parity states, in contrast to the only
previous work to consider decays of these states into the strange
final states $\Lambda K$ and $\Sigma K$.
\end{abstract}  
\pacs{13.30.Eg,12.40Qq,14.20Gk}
\newpage 
\def\slash#1{#1 \hskip -0.5em / }  

\section{Introduction}

The strange quark plays a unique role in particle and nuclear
physics. It is not quite light enough for expansions based on chiral
symmetry to work as well as they do in the case of the up and down
quarks, nor is it heavy enough for it to be safely treated in the
recently developed heavy quark effective theory. Since its
constituent mass is close to the typical scale of soft QCD
interactions, expansions in the ratio of its mass to this energy scale
may have convergence problems. The strange quark has also been very
important in the development of the standard model, as hadrons
containing strange quarks were the first to manifest flavor-changing
neutral currents and CP violation.

This unique nature makes the strange quark and its hadrons the object
of many theoretical and experimental studies. As an example, one can
ask whether non-perturbative QCD is flavor-blind. Do hadrons
containing strange quarks behave essentially like hadrons containing
only non-strange quarks, apart from differences in quark masses? Other
important and topical questions include those about the presence of
strangeness in the nucleon, the properties of hypernuclei, and
strangeness production in relativistic heavy-ion collisions as a
signature of the quark-gluon plasma.

A number of experimental facilities are currently engaged in studies
of strange hadrons, or will be in the near future. The kaon beam at
Brookhaven National Laboratory will clearly provide some impetus, as
will the pion beam there, through kaon production experiments. Recent
experiments at Bonn and Mainz, and future experiments at GRAAL and
especially at TJNAF will also stimulate interest in strange
matter. Indeed, the first two experiments to be completed at TJNAF
were kaon electroproduction experiments~\cite{kexpts1}.

Understanding how strange hadrons couple to non-strange hadrons is one
of the primary goals of the research described above. This
understanding will also be required to interpret experiments which
search for the presence of strangeness in the nucleon, and which
produce strange hadrons in relativistic heavy-ion
collisions. Calculations of such couplings are therefore essential to
our effort to understand experiments involving strange hadrons.

Kaon electromagnetic production experiments at
TJNAF~\cite{Kexpts,E-89-024} can be thought of as producing nonstrange
baryons in the $s$ channel, which subsequently decay into a strange
baryon and a strange meson. Analysis of kaon production experiments
using this model will yield information about the couplings of these
resonances to final states like $\Lambda K$, $\Sigma K$, and through
the detection of three body final states~\cite{E-89-024}, to final
states involving excited strange baryons and mesons such as $\Lambda
K^*$, $\Sigma K^*$, $\Sigma(1385)K$, {\it etc.} Because of the
relatively high thresholds for these final states compared to $N\pi$
and $N\pi\pi$, it can be expected that these decays are a good way to
study the poorly understood excited negative-parity baryons, which in
our model have wave functions predominantly in the $N=3$ oscillator
band. Few of the many states predicted to be present by symmetric
quark models have been seen in $N\pi$ elastic and inelastic
scattering, and many of those that have been seen are tentative
states. Some of those states already seen have masses significantly
lighter than model predictions~\cite{FC,CI}. This is a relatively young
aspect of baryon spectroscopy which deserves attention, and we will show
that strangeness production experiments should find many of these new
states.

Predictions for the amplitudes for nonstrange baryons to decay into
strange final states will be useful when planning the analyses of
these experiments. Furthermore, through associated production into the
final state $\Sigma\pi K$, it may be possible to study the poorly
understood strange baryon $\Lambda(1405)$ and its spin partner
$\Lambda(1520)$, which decay to $\Sigma\pi$. The nature of these two
states is of fundamental importance to the understanding of the
interquark potential. There have been suggestions in the literature
(for a brief review see Ref.~\cite{PDGDalitz}) that the difficulty
encountered in fitting the mass of the $\Lambda(1405)$ in quark
potential models~\cite{spectrum,CI} can be explained if this state is
a $\bar{K}N$ bound state. The problem in potential models is
that the $J^P=\frac{3}{2}^-$ state $\Lambda(1520)$ is predicted to be
essentially degenerate with its $J^P=\frac{1}{2}^-$ spin partner state
$\Lambda(1405)$. Spin-orbit interactions can lift this degeneracy, but
other aspects of the baryon spectrum rule out spin-orbit interactions
of the strength required to fit $\Lambda(1520)-\Lambda(1405)$. A good
way to resolve the controversy about the nature of this state is to
examine its strong~\cite{KI,FC} and electromagnetic decays~\cite{Koniuk}
with the assumption that it is a conventional three quark state; here we
search for one or more decays of a nucleon excited state
to $\Lambda(1405)K$ and $\Lambda(1520)K$ which give clear contrasting
predictions under this same assumption. If an experiment were to focus
on electromagnetic production of $\Lambda(1405)K$ and $\Lambda(1520)K$
through such an intermediate state, comparison with these predictions
could play an important role in solving this puzzle. Another
possibility, which we will examine in a later paper~\cite{CRstr2str},
is to examine the strong decays of higher-lying strange resonances
into the $\Lambda(1405)\pi$ and $\Lambda(1520)\pi$ channels.

Predictions for the decays of nonstrange baryons up to the $N=2$ band
to the $\Lambda K$ and $\Sigma K$ channels have been given by Koniuk
and Isgur~\cite{KI} in an elementary-meson emission model, where a
point-like kaon couples directly to the quarks in the initial
baryon. Forsyth and Cutkosky~\cite{FC} have also examined these final
state channels in a model based on a decay operator with the structure
${\bf S}\cdot (g_1{\bf P}_q+g_2{\bf P}_{\bar{q}})$, where ${\bf P}_q$
and ${\bf P}_{\bar{q}}$ are the momenta of the created quark and
antiquark, respectively, and ${\bf S}$ is their combined spin. Our
$^3P_0$ model~\cite{decays} corresponds to $g_1=0$; if $g_1$ and $g_2$
are allowed to depend on the state of the spectator quarks their model
allows for breaking of the usual spectator approximation. Forsyth and
Cutkosky's model predicts that there are a number of reasonably light
negative-parity states with wave functions predominantly in the $N=3$
band in a narrow energy range (from 2050-2300 MeV) which have large
amplitudes to decay to $\Sigma K$. As theirs is the only model of the
decays into these final states of $N=3$ band states, it is important
to verify this prediction.

In this work we provide predictions for the decay amplitudes into the
final states $\Lambda K$, $\Sigma K$, $\Lambda(1405) K$,
$\Lambda(1520) K$, $\Sigma(1385) K$, $\Lambda K^*$, and $\Sigma K^*$
of all states (seen and missing in $N\pi$) with wave functions
predominantly in the $N=1$ and $N=2$ bands, and also for several
low-lying states in higher bands, using the relativized model of
baryon decays based on the $^3P_0$ pair creation model of
Refs.~\cite{scwr} and~\cite{CR2}. Models of this kind are generally
more predictive than elementary-meson emission models, which usually
require a reduced matrix element to be fit to the decay of each type
[SU(6) multiplet] of initial baryon. The $^3P_0$ pair creation model
also properly takes into account the finite spatial extent of the
final meson. Another advantage is that we are able to extend this
model to include in the final state the excited strange baryons
$\Lambda(1405)$, $\Lambda(1520)$, and $\Sigma(1385)$, as well as the
excited meson $K^*$.

Model parameters are taken from our previous work and not adjusted;
wave functions are taken from the relativized model of Ref.~\cite{CI},
which describes all of the states considered here in a consistent
picture. In order to be in accord with the Particle Data Group
(PDG)~\cite{PDG} conventional definitions of decay widths, we have
determined the decay momentum using the central value of the PDG
quoted mass for resonances seen in $N\pi$, and the predicted mass from
Ref.~\cite{CI} for missing and undiscovered states. We have also
integrated over the line shape of the final state $K^*$, and
$\Sigma(1385)$ and $\Lambda(1405)$ baryons, with the phase space as
prescribed in the meson decay calculation of Ref.~\cite{KokI}; for
details of this procedure see Eq.~(8) of Ref.~\cite{CR2} (note that we
do not integrate over the narrow [16 MeV width] $\Lambda(1520)$
line shape). As a consequence there are states below the nominal
thresholds which have non-zero decay amplitudes.

In keeping with the convention of Ref.~\cite{CR2}, the phases of the
amplitudes are determined as follows. We quote the product
$A^{X\dag}_{Y K}A^X_{N\pi}/|A^X_{N\pi}|$ of the predicted decay
amplitude for $X\to Y K$ (where $Y$ is a ground state or excited state
hyperon and $K$ includes the $K^*$) and the phase of the decay
amplitude for $X\to N\pi$, the latter being unobservable in $N\pi$
elastic scattering (note factors of $+i$, conventionally suppressed in
quoting amplitudes for decays of negative parity baryons to $NM$ or
$N\gamma$, where $M$ has negative parity, do not affect this
product). This eliminates problems with (unphysical) sign conventions
for wave functions, and the relative signs of these products are then
predictions for the (physically significant) relative phases of the
contributions of states $X$ in the process $N\pi\to X\to YK$. The {\it
overall} phase of the $\Lambda K$ decay amplitudes quoted in the
Particle Data Group~\cite{PDG} cannot be determined experimentally and
so is fixed~\cite{Saxon} by choosing the sign of the $N\pi\to
S_{11}(1650)\to \Lambda K$ amplitude to be negative, as determined in
an $SU(6)_W\times O(3)$ analysis of the strong decays (see, for
example, Ref.~\cite{HeyLitch}). Similarly, the overall phase of the
$N\pi\to X\to \Sigma K$ decay amplitudes quoted in the Particle Data
Group~\cite{PDG} is fixed by comparison to the SU(3)$_f$ prediction
that the sign should be negative when $X$ is a $\Delta$
state~\cite{Candlin}. Since our calculation explicitly breaks
SU(3)$_f$, we fix the overall sign~\cite{Deans} by choosing the sign
of the amplitude for the low-lying state $X=\Delta(1950)F_{37}$, which
has a well measured amplitude, to be negative.  For the other final
state channels dealt with here it may be necessary to fix an
unmeasurable overall sign in the same way to compare with upcoming
analyses of new data.

For photo and electroproduction experiments at TJNAF and elsewhere it
may be useful to know the relative signs of the contributions of
states $X$ in the process $N\gamma\to X\to YK$. As the photocouplings
of Ref.~\cite{Cpc} are also quoted inclusive of the $N\pi$ sign,
$A^{X\dag}_{N\gamma}A^X_{N\pi}/|A^X_{N\pi}|$, then simply multiplying
the quoted photocouplings by the amplitudes quoted here will yield the
relative phases of the contributions of states $X$ in $N\gamma\to X\to
YK$.

\section{Results and Discussion}

Our results for decays into the $\Sigma K$, $\Sigma K^*$ and
$\Sigma(1385) K$ channels are given in Tables~\ref{NDle2SK}
and~\ref{NDge3SK}, and those for the $\Lambda K^*$, $\Lambda(1405) K$,
and $\Lambda(1520) K$ channels are given in Tables~\ref{NDle2LK}
and~\ref{NDge3LK}. We have listed the decay amplitudes into these
channels for each model state, which is also identified by its
assignment (if any) to a resonance from the analyses. The predictions
for the magnitude of the $N\pi$ decay amplitudes for each
state~\cite{scwr} and values for these magnitudes extracted from the
PDG~\cite{PDG} are also included for ease of identification of missing
resonances. All theoretical amplitudes are given with upper and lower
limits, along with the central value, in order to convey the
uncertainty in our results due to the uncertainty in the resonance's
mass.  These correspond to our predictions for the amplitudes for a
resonance whose mass is set to the upper and lower limits, and to the
central value, of the experimentally determined mass. For states as
yet unseen in the analyses of the data, we have adopted a `standard'
uncertainty in the mass of 150 MeV and used the model predictions for
the state's mass for the central value. If a state below the effective
threshold has been omitted from a table it is because our predictions
for all of its amplitudes are zero.

Figures~\ref{NgampiSK} to~\ref{DgampiSstarK} show the predictions of
the model of Ref.~\cite{CI} for the masses of excited $N^*$ and
$\Delta$ states below 2200 MeV, along with our predictions for the
square roots of the initial channel partial widths and the final
channel partial width for each state for the photoproduction reactions
$\gamma N\to X\to YK$ and for the pion production reactions $\pi N\to
X\to YK$. The final states $YK$ are those listed above, with the
exception of $\Sigma K^*$, for reasons discussed below. Photon partial
widths are calculated using the results of Ref.~\cite{Cpc}. When the
energy of the initial state in the center of momentum frame coincides
with the mass of a given resonance, the strength of the contribution
of that resonance will be proportional to the product of the initial
and final channel partial widths. We can estimate which states should
contribute strongly in a given energy region by comparing the products
of our predictions for the square roots of the initial and final
channel partial widths of states in that region. In a given production
process it should be possible to clearly separate nearby states in the
same partial wave when one of these states has this product small and
the other large. Model states in the figures which have well
established (three or four stars~\cite{PDG}) counterparts from the
analyses are distinguished from those which do not, in order to make
it simple to assess which new states may be seen in experiments of
this kind.

The amplitudes for decays into strange final states are generally
smaller than those into lighter nonstrange final states, as shown
below. However, it should still be possible to extract useful
information about intermediate nonstrange baryon resonances from
analyzing specific strange final states. In many strange channels only
a few higher mass states contribute with appreciable amplitudes, and
in several partial waves one or two states will dominate. This is due
in part to the higher thresholds in effect here, which allow these
channels to turn on in the mass region where new states are predicted
to be present by our model. This is to be contrasted with the
situation with nonstrange final states, where often low-lying states
with large amplitudes make extraction of information about higher mass
states with small amplitudes problematical.

The discussion below assumes the availability of polarization data for
these processes, so that partial wave analyses are possible. This is
automatic for reactions with final state $\Lambda$ baryons, as their
subsequent weak decays are self-analyzing. In addition, there are
plans for polarized beams and targets in experiments at Jefferson
Laboratory and elsewhere.

\subsection{$\Sigma K$ decays}

In general, channels involving $\Sigma$ or $\Sigma^*$ in the final
state will be difficult to analyze, as both $N$ and $\Delta$
resonances contribute to the cross section. This means that in many
instances one will encounter the difficulties associated with broad,
overlapping resonances. Nevertheless, our results show that it may
still be possible to confirm some weakly established resonances and
observe new states by analyzing these final states.

\begin{figure}
\epsfig{file=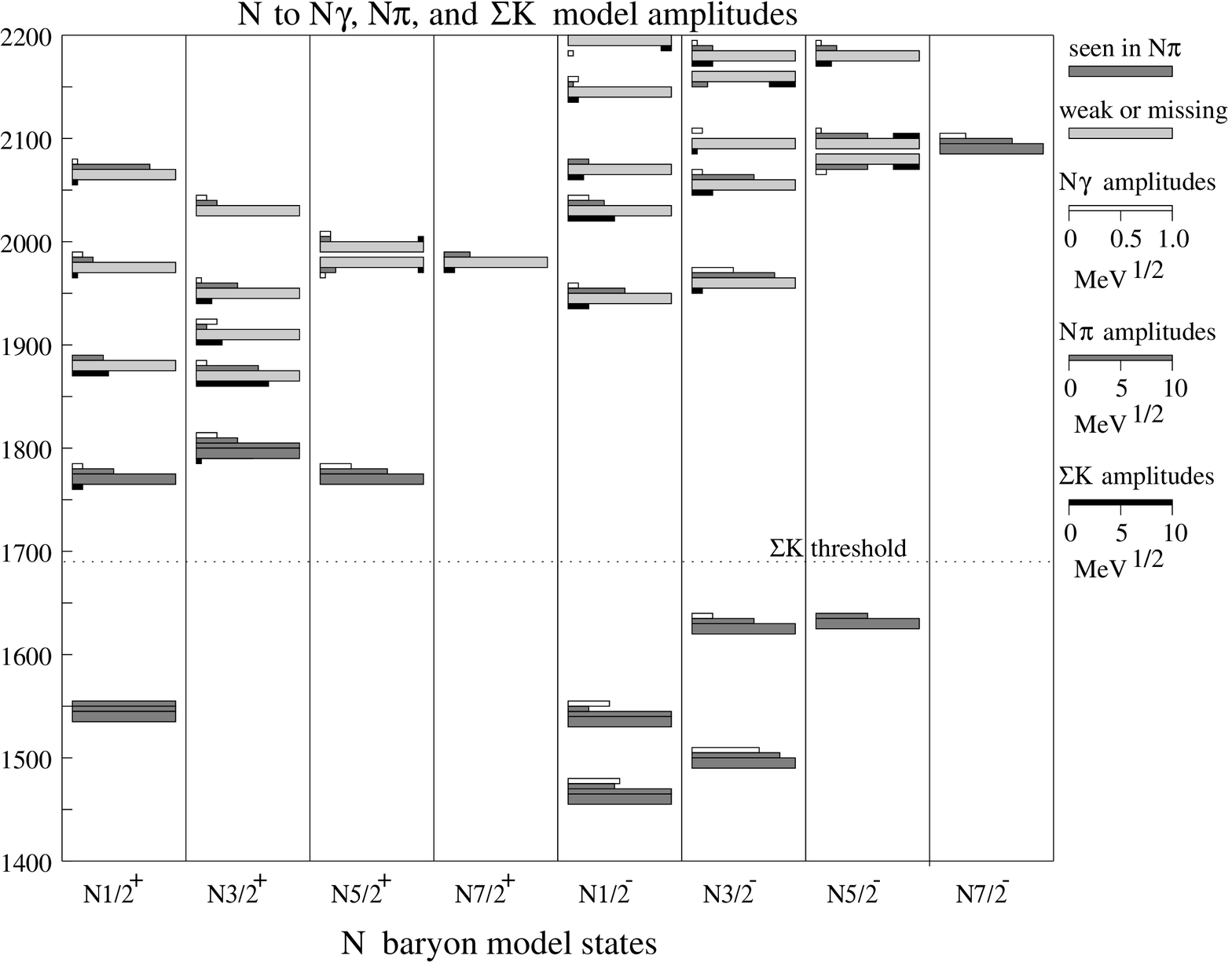,width=12cm,angle=-90}
\vskip 12pt
\caption{Mass predictions, $N\gamma$, $N\pi$, and $\Sigma K$ decay
amplitude predictions for nucleon resonances up to 2200 MeV, sorted
according to spin and parity. Heavy uniform-width bars show the
predicted masses of states with well established counterparts from
partial-wave analyses, light bars those of states which are weakly
established or missing. The length of the thin white bar gives our
prediction for each state's $N\gamma$ decay amplitude, that of the
thin grey bar gives our prediction for its $N\pi$ decay amplitude, and
that of the thin black bar gives our prediction for its $\Sigma K$
decay amplitude. States with significant amplitudes for
$N\gamma(N\pi)$ and $\Sigma K$ decays should contribute strongly to
the process $\gamma N (\pi N)\to\Sigma K$.}
\label{NgampiSK}
\end{figure}

\begin{figure}
\epsfig{file=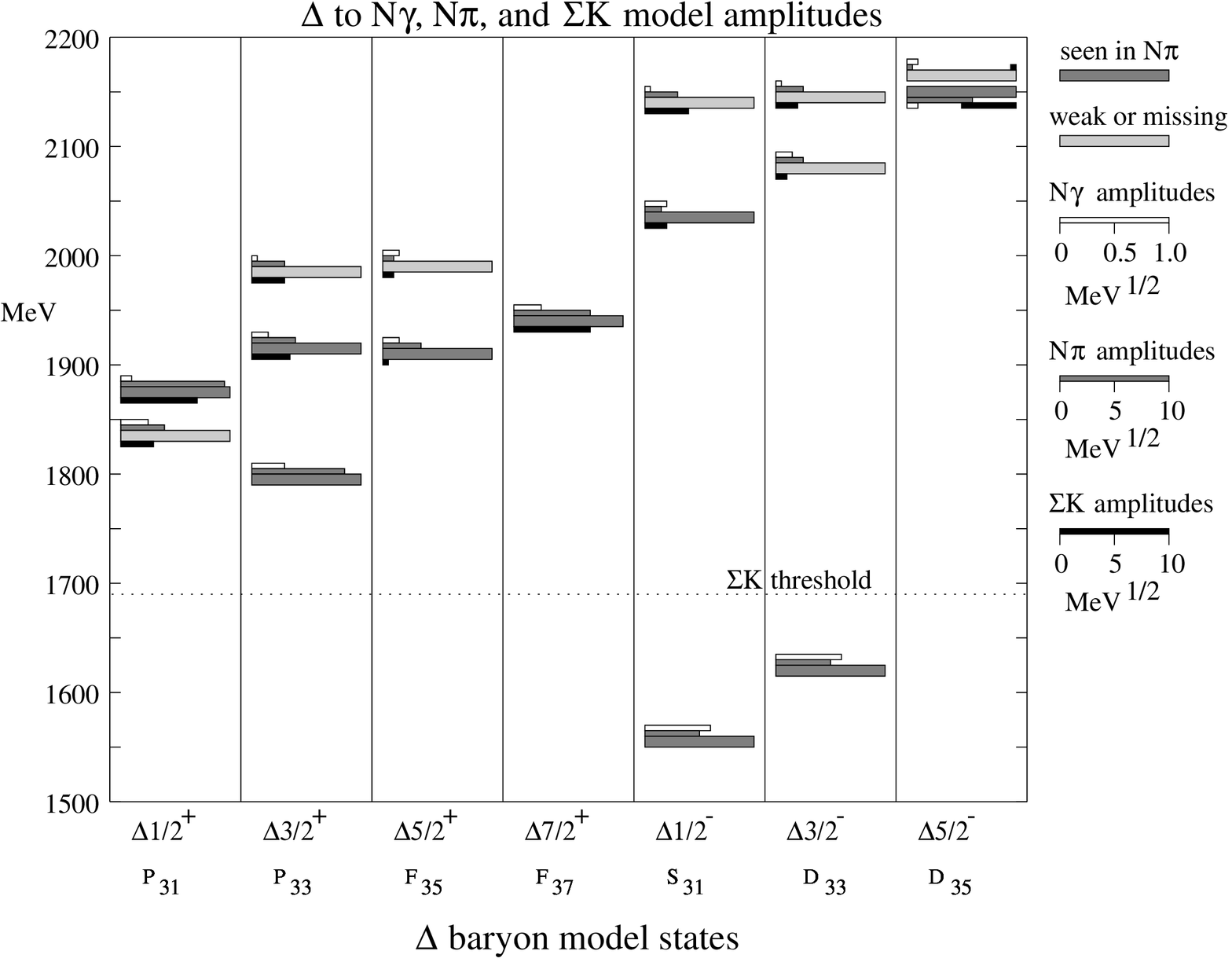,width=12cm,angle=-90}
\vskip 12pt
\caption{Mass predictions, $N\gamma$, $N\pi$, and $\Sigma K$ decay
amplitude predictions for $\Delta$ resonances up to 2200 MeV.
Notation as in Fig.~\protect{\ref{NgampiSK}}. States with significant
amplitudes for $N\gamma(N\pi)$ and $\Sigma K$ decays should contribute
strongly to the process $\gamma N (\pi N)\to\Sigma K$.}
\label{DgampiSK}
\end{figure}

The third columns of Tables~\ref{NDle2SK} and~\ref{NDge3SK} give our
predictions for the $\Sigma K$ decay amplitudes for nucleon and
$\Delta$ resonances with wave functions predominantly in the $N=1$ and
$2$ bands, and in higher bands, respectively. Our predictions for the
relative contributions of model states below 2200 MeV to
photoproduction and pion production of the $\Sigma K$ final state are
illustrated in Figs.~\ref{NgampiSK} and~\ref{DgampiSK}. The amplitudes
for this channel extracted from analyses are considerably less certain
than those for the $\Lambda K$ channel. However, there are a few
examples of states with substantial predicted amplitudes which have
also been seen with some certainty with this final state, such as
$\Delta(1950)F_{37}=[\Delta\frac{7}{2}^+]_1(1940)$ and
$\Delta(1920)P_{33}=[\Delta\frac{3}{2}^+]_3(1915)$ (the notation here
and in what follows is that states which have been seen in the
analyses are referred to by their PDG masses~\cite{PDG} and $N \pi$
partial wave, accompanied by our model state assignment). An
interesting discrepancy between our (substantial) prediction and the
extracted amplitudes exists for the state
$\Delta(1910)P_{31}=[\Delta\frac{1}{2}^+]_2(1875)$, for which an upper
limit only is quoted (see Table~\ref{NDle2SK} and the PDG~\cite{PDG}),
although older experiments admit the possibility of a larger
amplitude. This state is predicted to contribute strongly to both
$\gamma N\to \Sigma K$ and $\pi N\to \Sigma K$.

>From Figs.~\ref{NgampiSK} and~\ref{DgampiSK} we see that we predict
that a clear signal for the $N=2$ band missing positive-parity states
$N[\frac{1}{2}^+]_4(1880)$ and $\Delta[\frac{3}{2}^+]_4(1985)$ should
be present in the process $\pi N\to \Sigma K$. The $N=2$ band missing
state $N[\frac{3}{2}^+]_3(1910)$ should also be visible in $\gamma
N\to \Sigma K$. We also predict that the model states
$N[\frac{3}{2}^+]_2(1870)$ and $\Delta[\frac{1}{2}^+]_1(1835)$,
evidence for which is found in the multi-channel analysis of Manley
and Saleski~\cite{MANSA}, should contribute strongly to both photo-
and pion production of this final state. This suggests that an
analysis of these reactions may provide further evidence for these new
states.

>From Figs.~\ref{NgampiSK} and~\ref{DgampiSK} we see that several
low-lying negative-parity nucleon and $\Delta$ states with wave
functions predominantly in the $N=3$ band should contribute strongly
to photo- or pion production of the $\Sigma K$ final state. These
include the weakly established states
$N(2090)S_{11}=[N\frac{1}{2}^-]_3(1945)$,
$N(2200)D_{15}=[N\frac{5}{2}^-]_3(2095)$ (pion production), and
$\Delta(2150)S_{31}=\Delta[\frac{1}{2}^-]_3(2140)$, as well as the
three-star states $\Delta(1900)S_{31}=\Delta[\frac{1}{2}^-]_2(2035)$
(in photoproduction) and
$\Delta(1930)D_{35}=\Delta[\frac{5}{2}^-]_1(2155)$. Note that our
$\Sigma K$ amplitudes for these $\Delta$ states disagree with the
upper limits set by one analysis~\cite{Candlin}. Our results predict a
clear signal for the model state $N[\frac{1}{2}^-]_4(2030)$ in both
photo- and pion production of $\Sigma K$. The model state
$N[\frac{5}{2}^-]_2(2080)$ should also contribute strongly to $\pi
N\to \Sigma K$ and $\gamma N\to \Sigma K$, with a clear separation
from its nearby partner $N(2200)D_{15}=[N\frac{5}{2}^-]_3(2095)$ in
this partial wave in photoproduction.

Our predictions agree in sign (up to an overall sign which cannot be
determined experimentally) and largely in magnitude with those of
Koniuk and Isgur~\cite{KI} for the decays of states with wave functions
predominantly in the $N=1$ and $N=2$ bands. In two cases [the missing
state $N[\frac{3}{2}^+]_2(1870)$ and
$\Delta(1910)P_{31}=[\Delta\frac{1}{2}^+]_2(1875)$] our predicted
amplitudes are large, and substantially larger than those of Koniuk
and Isgur. Our model does not, however, confirm Forsyth and Cutkosky's
prediction~\cite{FC} that there are many light negative-parity states
with masses between 2050 and 2300 MeV which have large amplitudes to
decay to $\Sigma K$. For those $N=3$ band states for which they
predict large (up to 40 MeV) $\Sigma K$ widths, we have amplitudes
which are at most 2.4 MeV$^{\frac{1}{2}}$ in magnitude.

\subsection{$\Lambda K$ decays}

\begin{figure}
\epsfig{file=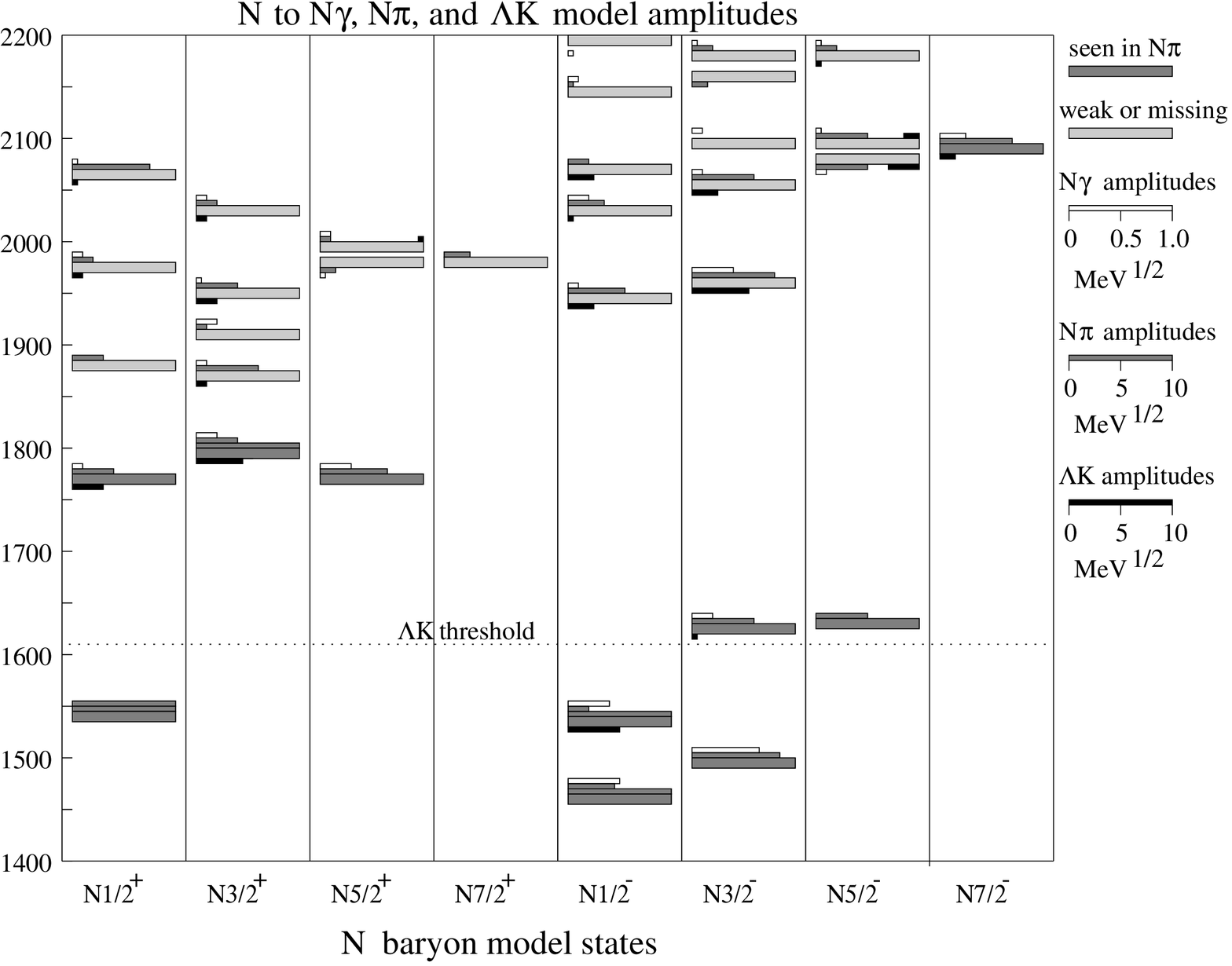,width=12cm,angle=-90}
\vskip 12pt
\caption{Mass predictions, $N\gamma$, $N\pi$, and $\Lambda K$ decay
amplitude predictions for nucleon resonances up to 2200 MeV.  Notation
as in Fig.~\protect{\ref{NgampiSK}}. States with significant
amplitudes for $N\gamma(N\pi)$ and $\Lambda K$ decays should
contribute strongly to the process $\gamma N (\pi N)\to\Lambda K$.}
\label{NgampiLK}
\end{figure}

The isospin selectivity of this final state means that only $N^*$
resonances (as opposed to $\Delta$ resonances) can be intermediate
states, which will simplify the analysis that will be required. Our
results for this final state are shown in the first columns of
Tables~\ref{NDle2LK} and~\ref{NDge3LK}, and our predictions for the
relative contributions of model states below 2200 MeV to photo- and
pion production of this final state are illustrated in
Figure~\ref{NgampiLK}.

The signs and magnitudes of the predicted $\Lambda K$ amplitudes are
in good agreement with the amplitudes extracted from the analyses
(largely in Table~\ref{NDle2LK}) for well determined states for which
there are substantial amplitudes. This gives us confidence that our
predictions are reliable. For example, our result for the
experimentally well determined $\Lambda K$ decay amplitude for
$N(1650)S_{11}=[N\frac{1}{2}^-]_2(1535)$ essentially agrees with that
of Forsyth and Cutkosky~\cite{FC}, and is a little larger than Koniuk
and Isgur's prediction~\cite{KI}; all are within errors of the
amplitude extracted from the analyses. Our results for states with
wave functions predominantly in the $N=1$ and $N=2$ bands largely
agree in both sign and in magnitude with those of Koniuk and
Isgur~\cite{KI}; there are just two states where the predicted signs
differ, and these have small predicted amplitudes. With the possible
exception of the model state $N[\frac{3}{2}^+]_4(1950)$ in pion
production, our model predicts that none of the missing
positive-parity nucleon states in the $N=2$ band has a substantial
$\Lambda K$ decay amplitude (see Fig.~\ref{NgampiLK}).

>From Fig.~\ref{NgampiLK} and Table~\ref{NDge3LK} we see that a
$\Lambda K$ experiment should show clear signals for several
relatively light negative-parity states with wave functions
predominantly in the $N=3$ band. The two-star state
$N(2080)D_{13}=[N\frac{3}{2}^-]_3(1960)$ should be clearly confirmed
with a precision $N\gamma$ or $N\pi\to \Lambda K$ experiment as it
should dominate its partial wave; the amplitudes quoted in the
Particle Data Group (PDG)~\cite{PDG} for this decay are smaller than
our prediction, but are without error estimates. The nearby model
state $N[\frac{3}{2}^-]_4(2055)$ should also contribute strongly to
$N\pi\to \Lambda K$. The model state $N[\frac{5}{2}^-]_2(2080)$ and
the nearby weak state $N(2200)D_{15}=N[\frac{5}{2}^-]_3(2095)$ should
dominate their partial wave in pion production of this final state,
with the former once again being clearly separated from its partner in
photoproduction.  The weakly established state
$N(2090)S_{11}=[N\frac{1}{2}^-]_3(1945)$ should also be visible in
both processes. Our predictions for these decays appear to be
substantially larger than those of Forsyth and Cutkosky~\cite{FC}, who
predict few appreciable $\Lambda K$ widths for states in the $N=3$
band.

\subsection{$\Lambda K^*$, $\Lambda(1405)K$ and $\Lambda(1520)K$
decays}

\begin{figure}
\epsfig{file=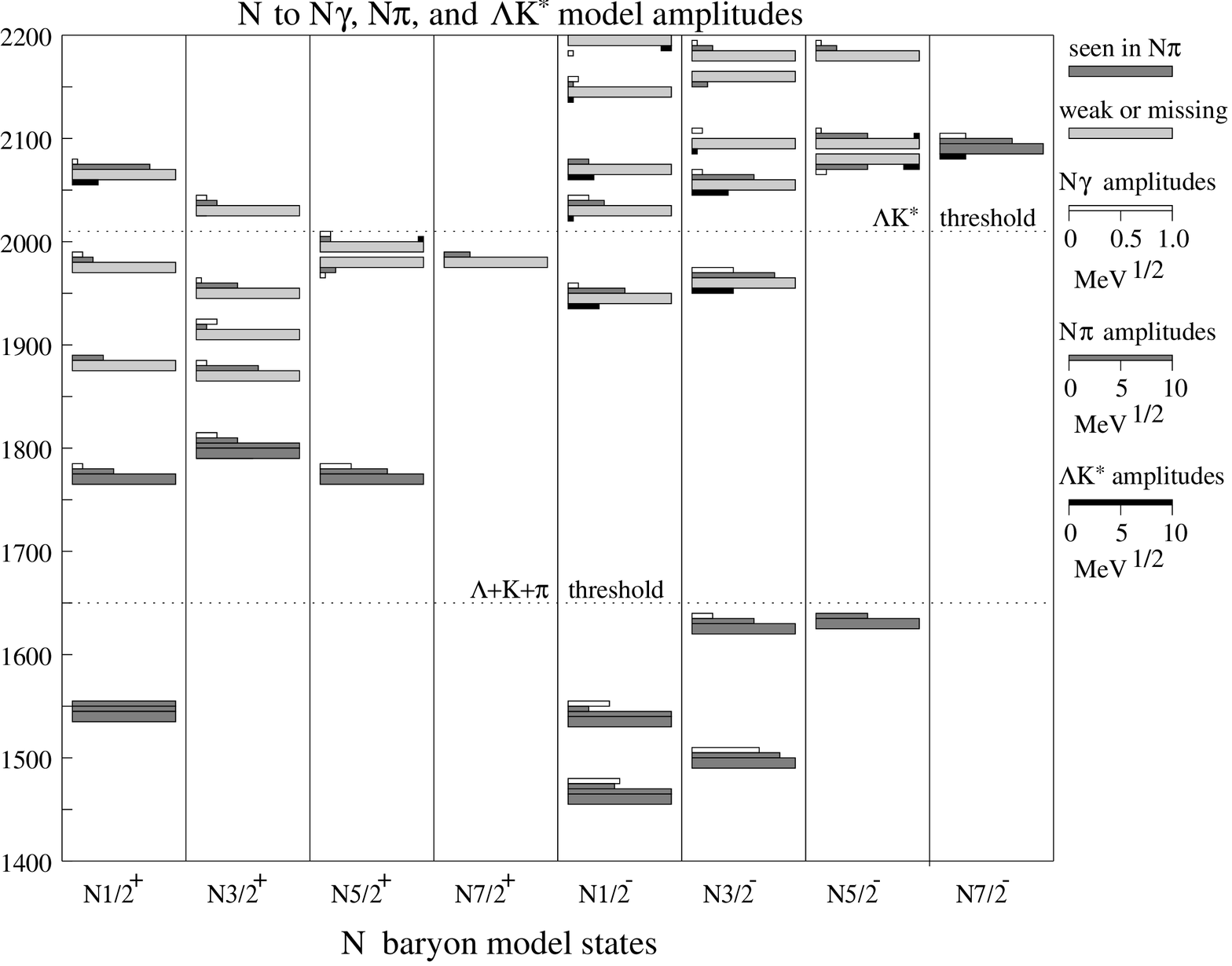,width=12cm,angle=-90}
\vskip 12pt
\caption{Mass predictions, $N\gamma$, $N\pi$, and $\Lambda K^*$ decay
amplitude predictions for nucleon resonances up to 2200 MeV.  Notation
as in Fig.~\protect{\ref{NgampiSK}}. States with significant
amplitudes for $N\gamma(N\pi)$ and $\Lambda K^*$ decays should
contribute strongly to the process $\gamma N (\pi N)\to\Lambda K^*$.}
\label{NgampiLKstar}
\end{figure}

\begin{figure}
\epsfig{file=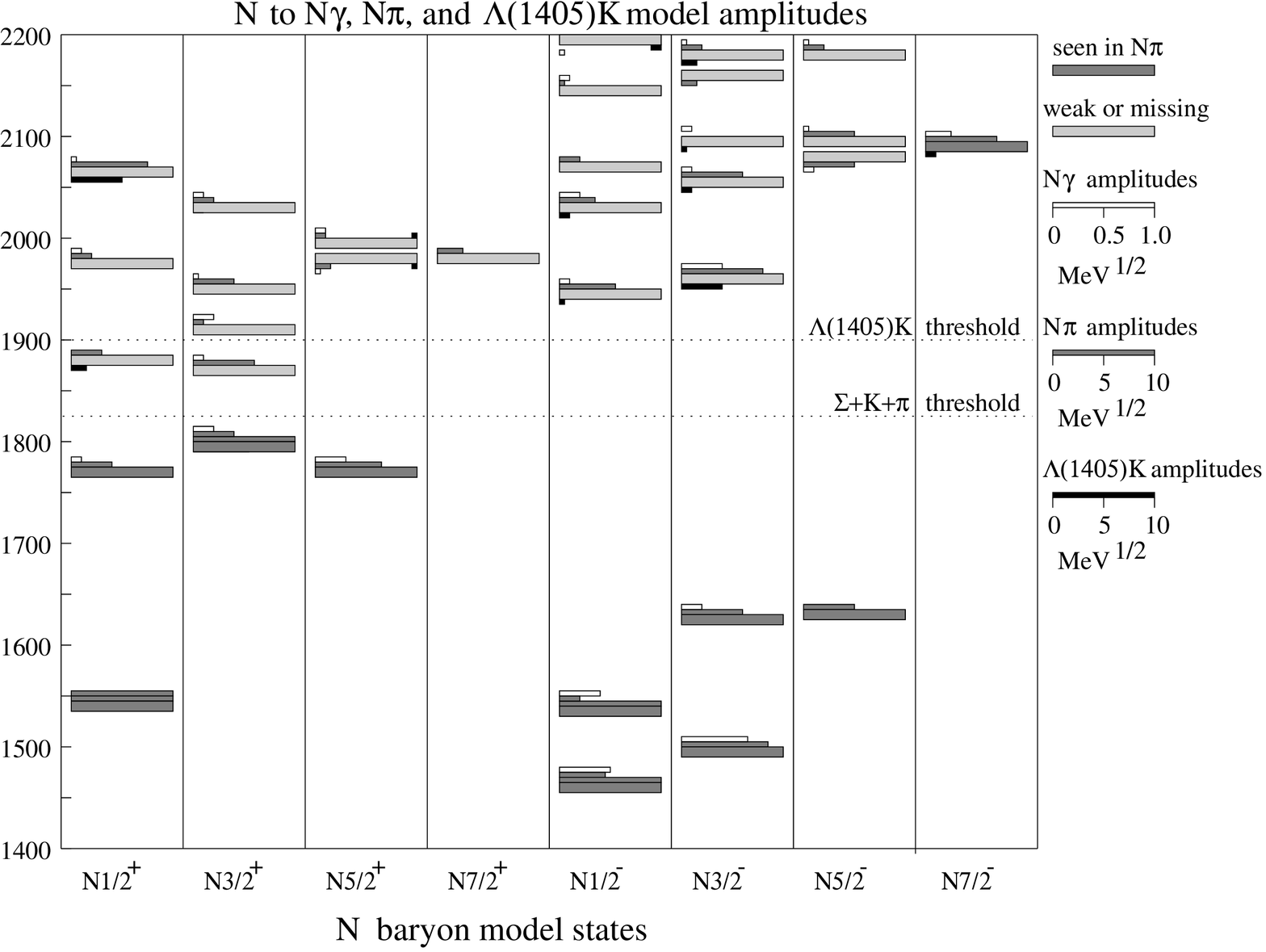,width=12cm,angle=-90}
\vskip 12pt
\caption{Mass predictions, $N\gamma$, $N\pi$, and $\Lambda(1405) K$
decay amplitude predictions for nucleon resonances up to 2200 MeV.
Notation as in Fig.~\protect{\ref{NgampiSK}}. States with significant
amplitudes for $N\gamma(N\pi)$ and $\Lambda(1405) K$ decays should
contribute strongly to the process $\gamma N (\pi N)\to\Lambda(1405)
K$.}
\label{NgampiL1405K}
\end{figure}

\begin{figure}
\epsfig{file=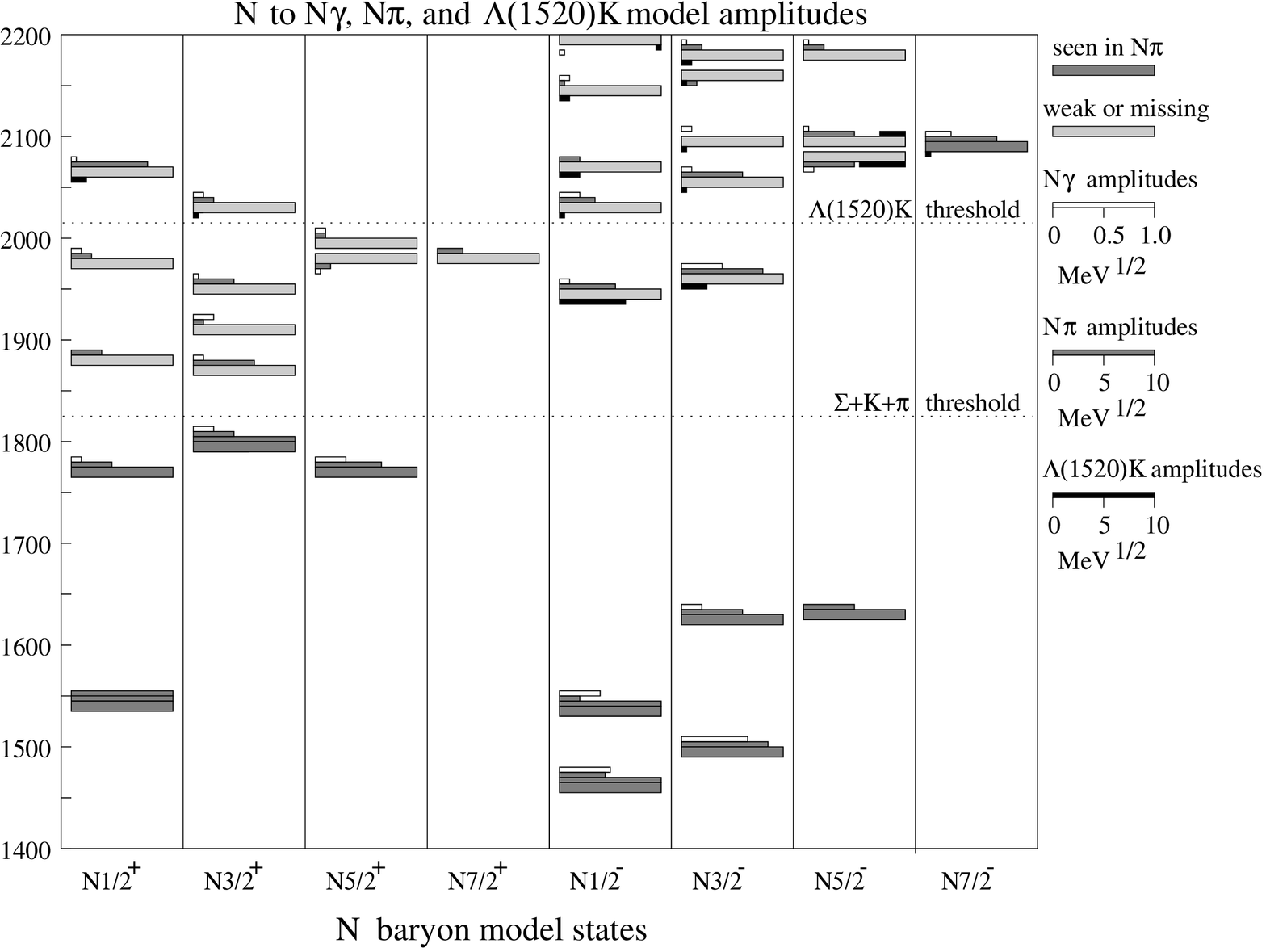,width=12cm,angle=-90}
\vskip 12pt
\caption{Mass predictions, $N\gamma$, $N\pi$, and $\Lambda(1520) K$
decay amplitude predictions for nucleon resonances up to 2200 MeV.
Notation as in Fig.~\protect{\ref{NgampiSK}}. States with significant
amplitudes for $N\gamma(N\pi)$ and $\Lambda(1520) K$ decays should
contribute strongly to the process $\gamma N (\pi N)\to\Lambda(1520)
K$.}
\label{NgampiL1520K}
\end{figure}

Our results for these final states are shown in Tables~\ref{NDle2LK}
and~\ref{NDge3LK}, and our predictions for the relative contributions
of model states below 2200 MeV to photo- and pion production of this
final state are illustrated in
Figs.~\ref{NgampiLKstar},~\ref{NgampiL1405K}
and~\ref{NgampiL1520K}. The figures show that, with the possible
exception of the missing state $N[\frac{1}{2}^+]_4(1880)$ which may be
visible in pion production of $\Lambda(1405)K$, our model predicts no
substantial contributions to these channels for any states in the
$N=2$ band.

Our predictions for $\Lambda K^*$ decays of selected states in higher
bands are shown in Table~\ref{NDge3LK} and in
Figure~\ref{NgampiLKstar}. Several low-lying weakly-established and
predicted negative-parity nucleon resonances should contribute
strongly to photo- and pion production of this final state; these
largely correspond to states mentioned above as important in $\Lambda
K$ production. In addition, our results show it may be possible to see
the well established state $N(2190)G_{17}=[N\frac{7}{2}^-]_1(2090)$ in
both production experiments. The tentative state
$N(2100)P_{11}=[N\frac{1}{2}^+]_6(2065)$ should also contribute
strongly to $\pi N\to \Lambda K^*$ (without interference from the
nearby model state $N[\frac{1}{2}^+]_7(2210)$, which decouples from
both $N\pi$ and $N\gamma$ [see Tables~\ref{NDge3SK} and~\ref{NDge3LK}]).

Our results for the $\Lambda(1405)K$ and $\Lambda(1520)K$ channels are
quite interesting, with substantial widths for several low-lying
negative-parity and two positive-parity states (see
Table~\ref{NDge3LK} and Figs.~\ref{NgampiL1405K}
and~\ref{NgampiL1520K}). The weakly established state
$N(2090)S_{11}=[N\frac{1}{2}^-]_3(1945)$ and the model state
$[N\frac{5}{2}^-]_2(2080)$ should be easily visible in the
$\Lambda(1520)K$ channel in both photo- and pion production
experiments, and the two-star $N(2080)D_{13}=[N\frac{3}{2}^-]_3(1960)$
state should contribute strongly to $\Lambda(1405)K$ and
$\Lambda(1520)K$ final channels in both production processes. The
tentative state $N(2100)P_{11}=[N\frac{1}{2}^+]_6(2065)$ has a large
predicted effect in $\Lambda(1405)K$ production, and the weak state
state $N(2200)D_{15}=N[\frac{5}{2}^-]_3(2095)$ and the model state
$N[\frac{7}{2}^+]_2(2390)$ (see Table~\ref{NDge3LK}) should be
prominent in the pion production of $\Lambda(1520)K$. Once again, only
the lighter $[N\frac{5}{2}^-]_2(2080)$ state should be visible in
$\gamma N \to \Lambda(1520)K$.

Just as importantly, we see that there are indeed clear predictions of
our model for the relative strengths of the decays of the lightest of
those states which decay strongly into $\Lambda(1405)K$ and
$\Lambda(1520)K$. Both of the weakly established states
$N(2090)S_{11}=[N\frac{1}{2}^-]_3(1945)$ and
$N(2200)D_{15}=[N\frac{5}{2}^-]_3(2095)$ are predicted to contribute
strongly to $\Lambda(1520)K$ production, but not to $\Lambda(1405)K$
production. The opposite is true of the weak $P_{11}$ state
$N(2100)=[N\frac{1}{2}^+]_6(2065)$. The $D_{13}$ state
$N(2080)=[N\frac{3}{2}^-]_3(1960)$ is predicted to appear with roughly
equal strength in both production of $\Lambda(1405)K$ and
$\Lambda(1520)K$. In addition, intermediate states in other partial
waves are predicted to contribute little to the production of these
final states. Furthermore, we predict that none of the well established
states in this mass region should couple strongly to
$\Lambda(1405)K$ or $\Lambda(1520)K$. A photo- or pion production
experiment in the region 2000-2300 MeV which is able to identify the
overall spin and parity of the final state and reconstruct these two
final baryons, although difficult, would be able to test these
predictions of our three-quark model for the relative sizes of these
decay amplitudes, and possibly resolve the issue of the nature of the
$\Lambda(1405)$.

\subsection{$\Sigma K^*$ and $\Sigma(1385)K$ decays}

\begin{figure}
\epsfig{file=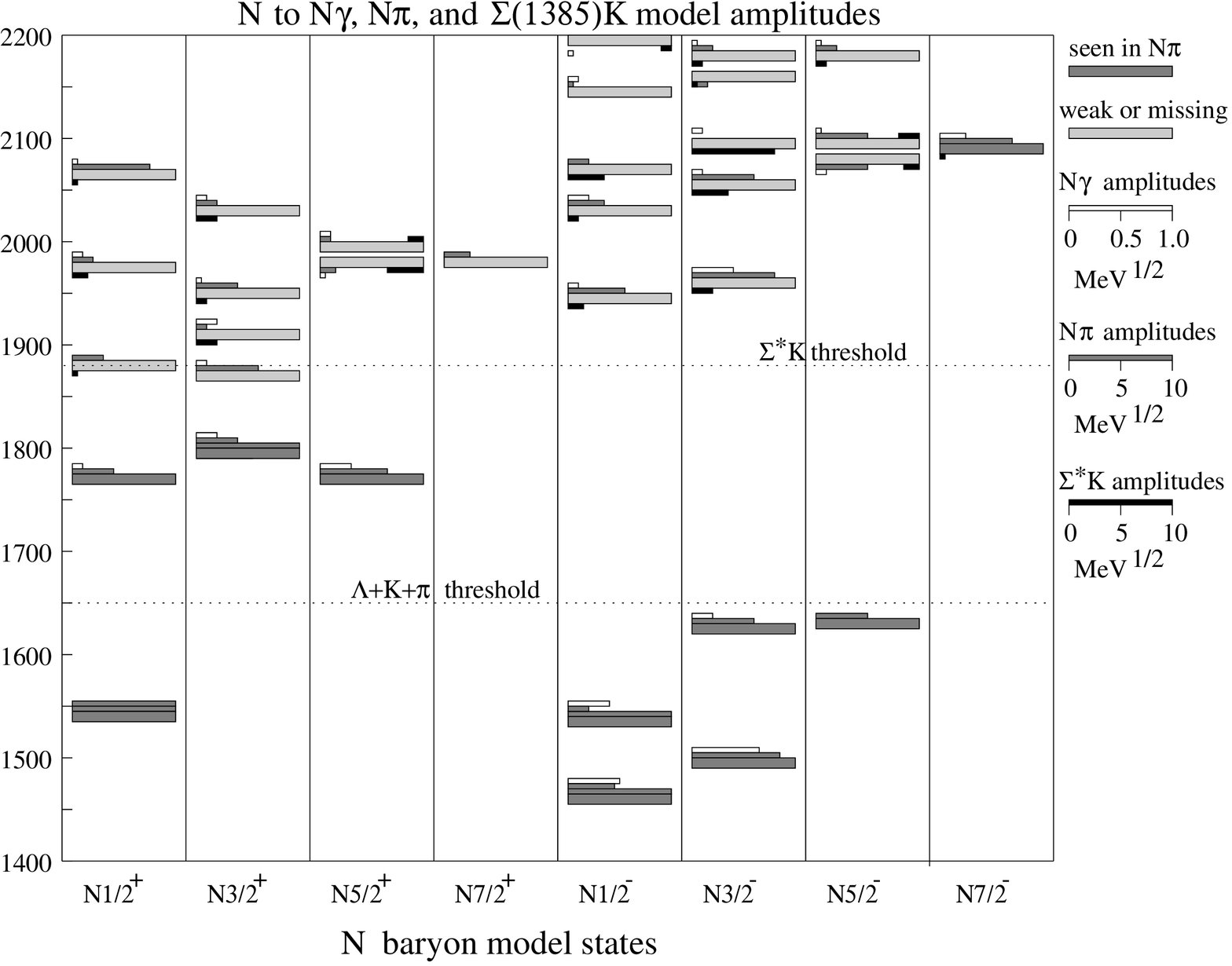,width=12cm,angle=-90}
\vskip 12pt
\caption{Mass predictions, $N\gamma$, $N\pi$, and $\Sigma(1385) K$
decay amplitude predictions for nucleon resonances up to 2200 MeV.
Notation as in Fig.~\protect{\ref{NgampiSK}}. States with significant
amplitudes for $N\gamma(N\pi)$ and $\Sigma(1385) K$ decays should
contribute strongly to the process $\gamma N (\pi N)\to\Sigma(1385)
K$.}
\label{NgampiSstarK}
\end{figure}

\begin{figure}
\epsfig{file=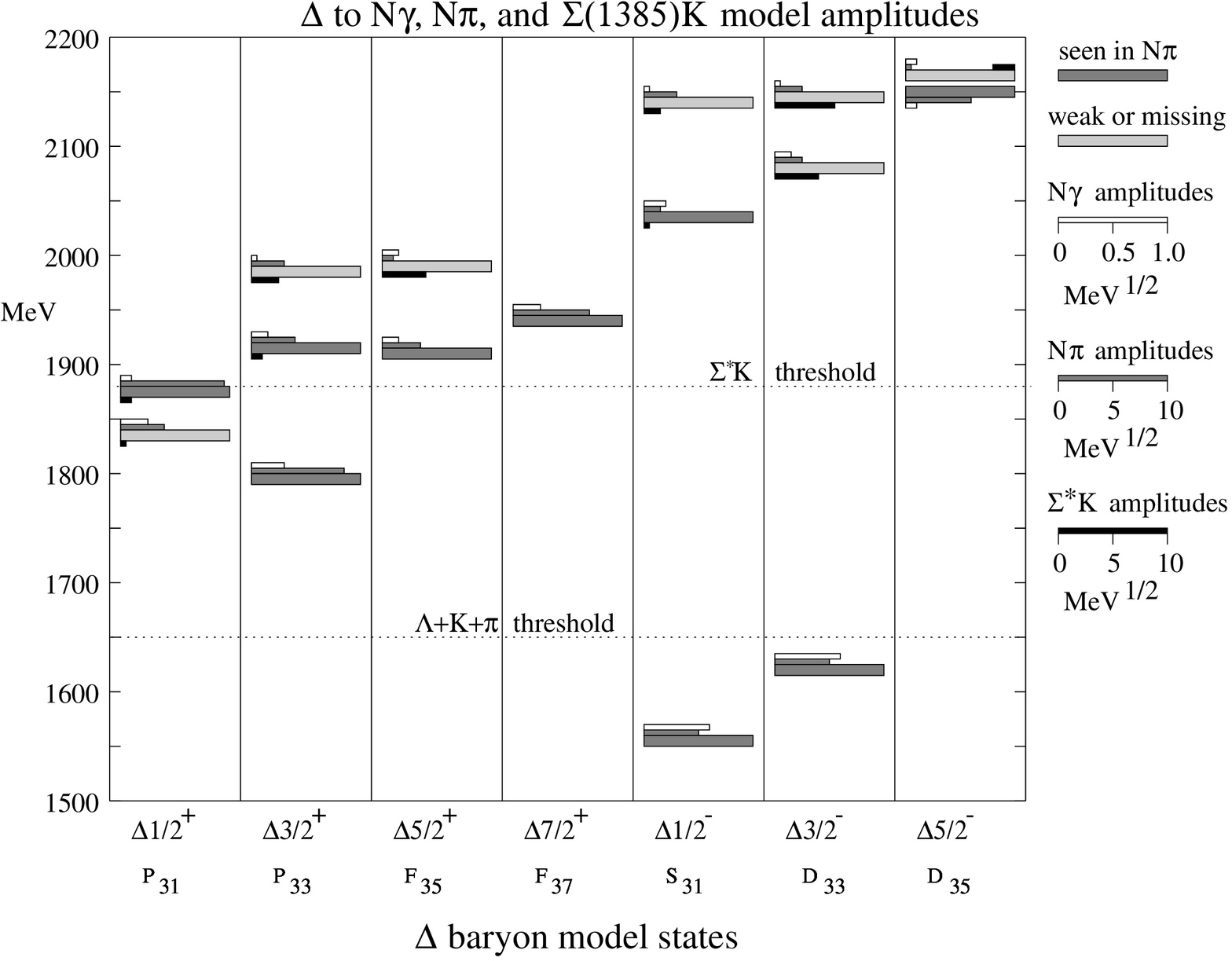,width=12cm,angle=-90}
\vskip 12pt
\caption{Mass predictions, $N\gamma$, $N\pi$, and $\Sigma(1385) K$
decay amplitude predictions for $\Delta$ resonances up to 2200 MeV.
Notation as in Fig.~\protect{\ref{NgampiSK}}. States with significant
amplitudes for $N\gamma(N\pi)$ and $\Sigma(1385) K$ decays should
contribute strongly to the process $\gamma N (\pi N)\to\Sigma(1385)
K$.}
\label{DgampiSstarK}
\end{figure}

>From the amplitudes in Table~\ref{NDle2SK} and from
Figs.~\ref{NgampiSstarK} and~\ref{DgampiSstarK} we see that it may be
possible to discover the $N=2$ band missing states
$N[\frac{5}{2}^+]_2(1980)$ and $\Delta[\frac{3}{2}^+]_4(1985)$ and
confirm the state $\Delta(2000)F_{35}=\Delta[\frac{5}{2}^+]_2(1990)$
in a $\Sigma(1385)K$ production experiment. Contributions to $\Sigma
K^*$ production from these light states are weak, which is due in part
to the higher nominal threshold.

Our results for decays of higher-lying states into these channels are
shown in Table~\ref{NDge3SK} and Figs.~\ref{NgampiSstarK}
and~\ref{DgampiSstarK}. We have not included a figure for the $\Sigma
K^*$ channel, as from Table~\ref{NDge3SK} we see that few low-lying
negative-parity predicted states should contribute strongly to
production of $\Sigma K^*$. These include $N[\frac{1}{2}^-]_5(2070)$
(in pion production) and $\Delta[\frac{3}{2}^-]_3(2145)$, and the
weakly established state
$\Delta(2150)S_{31}=\Delta[\frac{1}{2}^-]_3(2140)$. The first two of
these states are predicted to also contribute strongly to
$\Sigma(1385)K$ production. From Fig.~\ref{NgampiSstarK} we see that
the dominant contributions to pion production of $\Sigma(1385)K$ in
their partial wave should come from the weak state
$N(2080)D_{13}=[N\frac{3}{2}^-]_3(1960)$ and the predicted state
$N[\frac{3}{2}^-]_4(2055)$, whereas the dominant contribution to
photoproduction in this partial wave should come from the nearby state
$N[\frac{3}{2}^-]_5(2095)$. It may also be possible to confirm the
tentative state $\Delta(1940)D_{33}=\Delta[\frac{3}{2}^-]_2(2080)$ in
a $\Sigma(1385)K$ production experiment.

\section{Conclusions}

Our predictions for the $\Lambda K$ and $\Sigma K$ decays of low-lying
nonstrange baryons are similar to those given by Koniuk and Isgur, who
study states with wave functions predominantly in the $N=1$ and
$N=2$ bands. These results compare favorably in both sign and
magnitude with those amplitudes reliably determined from existing
data, which are largely to $\Lambda K$. There is a clear contrast
between our results for negative-parity states with wave functions
predominantly in the $N=3$ band and those of Forsyth and Cutkosky; we
predict more substantial $\Lambda K$ amplitudes, and do not confirm
their prediction of a narrow band of states with large amplitudes to
decay to $\Sigma K$.

If we consider an $s$-channel picture of the pion and electromagnetic
production of strange baryons and mesons, and assume a conventional
three-quark structure for the $\Lambda(1405)$, we see that the
$\Lambda(1405)K$ final state will be produced when $\sqrt{s}$ is of
the order of the mass of the intermediate states found here to have
appreciable couplings to this channel, roughly 2000--2300 MeV. It is
in this mass region also that the final state $\Lambda(1520)K$ will be
produced. Although the production amplitude in either case will be a
coherent sum of the amplitudes through a few intermediate states, it
should be possible to confirm or rule out a three-quark structure for
the $\Lambda(1405)$ by studying these channels and comparing to our
clear predictions for the relative sizes (and phases) of the
amplitudes for decays into these states.

Our results also show that several missing and undiscovered states
have substantial amplitudes to decay to strange final states, so that
if kaon electromagnetic and pion production experiments were to
focus on the region of 1800-2300 MeV, a careful analysis of the
results would be likely to discover many new baryon states and provide
much needed information about the parameters of states weakly
established in other channels. As most of these states are
negative-parity states with wave functions predominantly in the $N=3$
band, partially due to the relatively high thresholds for these final
states, the spectrum of many such states may be determined
conclusively for the first time in a strangeness production
experiment.

\section{Acknowledgements}
The authors would like to acknowledge stimulating discussions of many
of the issues addressed here with Professor Jim Napolitano. This work
was supported in part by the the Florida State University
Supercomputer Computations Research Institute which is partially
funded by the Department of Energy through Contract DE-FC05-85ER250000
(SC); the U.S. Department of Energy under Contract
No. DE-FG05-86ER40273 (SC); the National Science Foundation through
Grant No.\ PHY-9457892 (WR); the U.S.  Department of Energy under
Contract No.\ DE-AC05-84ER40150 (WR); and by the U.S. Department of
Energy under Contract No.\ DE-FG02-97ER41028 (WR).

\squeezetable

\begin{table}
\caption{Results for $N$ and $\Delta$ states in the $N=1$ and $N=2$
bands in the $\Sigma K$, $\Sigma K^*$, and $\Sigma(1385) K$
channels. $N\pi$ amplitudes from 
Ref.~\protect\cite{scwr}
are included to explain our assignments of the model states to
resonances. Notation for model states is $[J^P]_n$(mass[MeV]), 
where $J^P$ is the spin/parity of the state and $n$ its
principal quantum number. The first row gives our model results, while
the second row lists the available $N\pi$ and $\Sigma K$ amplitudes
from the partial-wave analyses, as well as the Particle Data Group
(PDG) name for the state, its $N\pi$ partial wave, and its PDG star
rating. Light states with zero amplitudes are omitted from the
table. Signs are omitted from experimental amplitudes where they are
not determined; with the exception of the $N\pi$ amplitudes (where we
do not quote predicted signs) we omit positive signs from amplitudes
predicted to be positive.}
\label{NDle2SK}
\begin{tabular}{lrrrrrrrrr}
 State & $N \pi$ & $\Sigma K$ & $\Sigma K^*$ & $\Sigma K^*$ & $\Sigma
  K^*$ & $\sqrt{\Gamma_{\Sigma K^*}}$ & $\Sigma(1385) K$ &
  $\Sigma(1385) K$ & $\sqrt{\Gamma_{\Sigma(1385) K}}$ \\ \tableline

 $[N\textstyle{1\over 2}^-]_2(1535)$ & 12.2 $\pm $ 0.8 \\

   $N(1650)S_{11}$**** & $11.5 \pm $ 1.3 & $\simeq$ 2.7 $\pm $ 1.8 &
   & & & & & & \\

 $[N\textstyle{3\over 2}^-]_2(1625)$ & 5.8 $\pm $ 0.6 & 0.0 $^{+
  0.3}_{- 0.0}$ \\
 
   $N(1700)D_{13}$*** & $3.2 \pm $ 1.1 & $<$ 0.5 & & & & & & & \\
 
 $[N\textstyle{5\over 2}^-]_1(1630)$ & 5.3 $\pm $ 0.1 \\
 
   $N(1675)D_{15}$**** & $8.2 \pm $ 0.9 & $<$ 0.1 & & & & & & & \\
 
 $[\Delta\textstyle{3\over 2}^-]_1(1620)$ & 4.9 $\pm $ 0.7 & 0.1 $^{+
  0.6}_{- 0.1}$ \\
 
   $\Delta(1700)D_{33}$**** & $6.7 \pm $ 1.6 & $\simeq$ 0.2 $\pm $
   0.1 & & & & & & & \\
 
& & & $p_{\frac{1}{2}}$ & $p_{\frac{3}{2}}$ & & & $p$ \\ \cline{4-5}
\cline{8-8}

 $[N\textstyle{1\over 2}^+]_3(1770)$ & 4.2 $\pm $ 0.1 & $-1.1 ^{+
  1.1}_{- 0.9}$ \\
 
   $N(1710)P_{11}$*** & $4.7 \pm $ 1.8 & $\simeq - 1.1 \pm $ 1.4 &
   & & & & & & \\
 
 $[N\textstyle{1\over 2}^+]_4(1880)$ & 2.7 $^{+ 0.6}_{- 0.9}$ & 3.7
  $^{+ 1.2}_{- 2.4}$ & 0.0 $^{+ 0.4}_{- 0.0}$ & 0.0 $\pm $ 0.1 & & 0.0
  $^{+ 0.4}_{- 0.0}$ & 0.4 $^{+ 2.8}_{- 0.4}$ & & 0.4 $^{+ 2.8}_{-
  0.4}$ \\
 
 $[N\textstyle{1\over 2}^+]_5(1975)$ & 2.0 $_{- 0.3}^{+ 0.2}$ & 0.6
  $\pm $ 0.1 & 0.1 $^{+ 1.0}_{- 0.1}$ & 0.0 $^{+ 0.0}_{- 0.2}$ & & 0.1
  $^{+ 1.0}_{- 0.1}$ & 1.3 $^{+ 1.1}_{- 1.3}$ & & 1.3 $^{+ 1.1}_{-
  1.3}$ \\
 
& & & $p_{\frac{1}{2}}$ & $p_{\frac{3}{2}}$ & $f_{\frac{3}{2}}$ & &
$p$ & $f$ \\ \cline{4-6} \cline{8-9}

 $[N\textstyle{3\over 2}^+]_1(1795)$ & 14.1 $\pm $ 0.1 & $-0.3 \pm $
  0.3 \\
 
   $N(1720)P_{13}$**** & $4.7 \pm $ 1.1 & $\simeq$ 2.2 $\pm $ 1.1 &
   & & & & & & \\
 
 $[N\textstyle{3\over 2}^+]_2(1870)$ & 6.1 $_{- 1.2}^{+ 0.6}$ & 7.0
  $^{+ 2.5}_{- 4.9}$ & 0.0 $\pm $ 0.0 & 0.0 $^{+ 0.0}_{- 0.4}$ & 0.0
  $\pm $ 0.0 & 0.0 $^{+ 0.4}_{- 0.0}$ & $-0.2 ^{+ 0.2}_{- 2.0}$ & 0.0
  $^{+ 0.2}_{- 0.0}$ & 0.2 $^{+ 2.0}_{- 0.2}$ \\
 
   & $11.4\pm 1.6^{\rm a}$\\
 
 $[N\textstyle{3\over 2}^+]_3(1910)$ & 1.0 $\pm $ 0.1 & 2.5 $^{+
  0.8}_{- 1.3}$ & 0.1 $^{+ 0.8}_{- 0.1}$ & $-0.1 ^{+ 0.1}_{- 0.9}$ &
  0.0 $\pm $ 0.0 & 0.1 $^{+ 1.2}_{- 0.1}$ & $-1.9 ^{+ 1.9}_{- 7.3}$ &
  0.0 $^{+ 0.0}_{- 0.4}$ & 1.9 $^{+ 7.3}_{- 1.9}$ \\
 
 $[N\textstyle{3\over 2}^+]_4(1950)$ & 4.1 $_{- 0.7}^{+ 0.4}$ & 1.4
  $^{+ 0.3}_{- 0.6}$ & 0.1 $^{+ 1.0}_{- 0.1}$ & 0.1 $^{+ 1.0}_{- 0.1}$
  & 0.0 $\pm $ 0.0 & 0.1 $^{+ 1.4}_{- 0.1}$ & 1.1 $^{+ 1.6}_{- 1.0}$ &
  $-0.1 ^{+ 0.1}_{- 0.7}$ & 1.1 $^{+ 1.7}_{- 1.0}$ \\
 
 $[N\textstyle{3\over 2}^+]_5(2030)$ & 1.8 $\pm $ 0.2 & 0.0 $\pm $
  0.0 & 0.1 $^{+ 0.6}_{- 0.1}$ & 0.3 $^{+ 2.1}_{- 0.3}$ & 0.0 $\pm $
  0.0 & 0.3 $^{+ 2.1}_{- 0.3}$ & 2.2 $^{+ 1.0}_{- 1.9}$ & $-0.2 ^{+
  0.1}_{- 0.3}$ & 2.2 $^{+ 1.0}_{- 1.9}$ \\
  
& & & $f_{\frac{1}{2}}$ & $p_{\frac{3}{2}}$ & $f_{\frac{3}{2}}$ & &
$p$ & $f$ \\ \cline{4-6} \cline{8-9}

 $[N\textstyle{5\over 2}^+]_2(1980)$ & 1.3 $\pm $ 0.2 & 0.4 $\pm $
  0.3 & 0.0 $\pm $ 0.0 & 0.1 $^{+ 0.3}_{- 0.1}$ & 0.0 $\pm $ 0.1 & 0.1
  $^{+ 0.3}_{- 0.1}$ & $-3.6 ^{+ 2.5}_{- 3.0}$ & $-0.1 ^{+ 0.1}_{-
  0.3}$ & 3.6 $^{+ 3.0}_{- 2.5}$ \\
 
 $[N\textstyle{5\over 2}^+]_3(1995)$ & 0.9 $\pm $ 0.2 & $-0.6 ^{+
  0.4}_{- 0.6}$ & 0.0 $\pm $ 0.0 & $-0.2 ^{+ 0.2}_{- 2.4}$ & 0.0 $^{+
  0.1}_{- 0.0}$ & 0.2 $^{+ 2.4}_{- 0.2}$ & $-1.7 ^{+ 1.6}_{- 1.3}$ &
  0.2 $^{+ 0.7}_{- 0.2}$ & 1.7 $^{+ 1.4}_{- 1.6}$ \\
 
   $N(2000)F_{15}$** & 4.2 $\pm $ 1.8 & $\simeq$ 2.5 $\pm $ 2.2 & &
   & & & & & \\
 
& & & & & & & $f$ & $h$ \\ \cline{8-9}

 $[N\textstyle{7\over 2}^+]_1(2000)$ & 2.4 $\pm $ 0.4 & 1.1 $^{+
  0.7}_{- 0.5}$ & & & & 0.0 $\pm $ 0.0 & $-0.2 ^{+ 0.2}_{- 0.5}$ & 0.0
  $\pm $ 0.0 & 0.2 $^{+ 0.5}_{- 0.2}$ \\
 
   $N(1990)F_{17}$** & 4.6 $\pm $ 1.2 & $\simeq$ 2.9 $\pm $ 2.2 & &
   & & & & & \\
 
& & & $p_{\frac{1}{2}}$ & $p_{\frac{3}{2}}$ & & & $p$ \\ \cline{4-5}
\cline{8-8}

$[\Delta\textstyle{1\over 2}^+]_1(1835)$ & 3.9 $^{+ 0.4}_{- 0.7}$ &
  2.9 $^{+ 1.4}_{- 2.9}$ & 0.0 $^{+ 0.0}_{- 0.4}$ & 0.0 $^{+ 0.3}_{-
  0.0}$ & & 0.0 $^{+ 0.5}_{- 0.0}$ & $-0.3 ^{+ 0.3}_{- 5.5}$ & & 0.3
  $^{+ 5.5}_{- 0.3}$ \\
 
   $\Delta(1740)P_{31}$$^{\rm b}$ & $4.9\pm 1.3$ & & & & & & & & \\
 
$[\Delta\textstyle{1\over 2}^+]_2(1875)$ & 9.4 $\pm $ 0.4 & 6.9 $^{+
  0.6}_{- 0.7}$ & 0.0 $\pm $ 0.1 & 0.0 $\pm $ 0.0 & & 0.1 $\pm $ 0.1 &
  1.0 $^{+ 0.9}_{- 0.7}$ & & 1.0 $^{+ 0.9}_{- 0.7}$ \\
 
   $\Delta(1910)P_{31}$**** & 7.5 $\pm $ 1.5 & $<$ 1.0 & & & & & & & \\
 
& & & $p_{\frac{1}{2}}$ & $p_{\frac{3}{2}}$ & $f_{\frac{3}{2}}$ & &
$p$ & $f$ \\ \cline{4-6} \cline{8-9}

$[\Delta\textstyle{3\over 2}^+]_2(1795)$ & 8.7 $\pm $ 0.2 & 0.0 $^{+
  1.1}_{- 0.0}$ \\
 
   $\Delta(1600)P_{33}$*** & 7.8 $\pm $ 2.0 & $\simeq$ 1.1 $\pm $ 0.9
   & & & & & & & \\
 
$[\Delta\textstyle{3\over 2}^+]_3(1915)$ & 4.2 $\pm $ 0.3 & 3.3 $\pm
  $ 0.3 & 0.1 $\pm $ 0.1 & 0.1 $\pm $ 0.1 & 0.0 $\pm $ 0.0 & 0.1 $\pm
  $ 0.2 & 1.2 $^{+ 1.3}_{- 0.9}$ & $-0.1 \pm $ 0.0 & 1.2 $^{+ 1.3}_{-
  0.9}$ \\
 
   $\Delta(1920)P_{33}$*** & 5.3 $\pm $ 1.8 & $\simeq$+2.2 $\pm $ 1.2
   & & & & & & & \\
 
$[\Delta\textstyle{3\over 2}^+]_4(1985)$ & 3.3 $^{+ 0.8}_{- 1.1}$ &
  3.2 $^{+ 0.3}_{- 0.9}$ & $-0.2 ^{+ 0.2}_{- 1.8}$ & 0.1 $^{+ 1.0}_{-
  0.1}$ & 0.0 $^{+ 0.0}_{- 0.2}$ & 0.2 $^{+ 2.1}_{- 0.2}$ & 2.6 $^{+
  2.1}_{- 2.5}$ & 0.1 $^{+ 0.5}_{- 0.1}$ & 2.6 $^{+ 2.1}_{- 2.5}$ \\
 
& & & $f_{\frac{1}{2}}$ & $p_{\frac{3}{2}}$ & $f_{\frac{3}{2}}$ & &
$p$ & $f$ \\ \cline{4-6} \cline{8-9}

$[\Delta\textstyle{5\over 2}^+]_1(1910)$ & 3.4 $\pm $ 0.2 & 0.4 $\pm $
  0.1 & 0.0 $\pm $ 0.0 & 0.0 $\pm $ 0.1 & 0.0 $\pm $ 0.0 & 0.0 $\pm $
  0.1 & 0.1 $\pm $ 0.1 & 0.0 $\pm $ 0.0 & 0.2 $\pm $ 0.1 \\
 
   $\Delta(1750)F_{35}$$^{\rm c}$ & $2.0\pm 0.8$ \\
   $\Delta(1905)F_{35}$**** & 5.7 $\pm $ 1.6 & $\simeq - 0.9 \pm $
   0.3 & & & & & & & \\
 
$[\Delta\textstyle{5\over 2}^+]_2(1990)$ & 1.2 $\pm $ 0.4 & 0.2 $^{+
  0.3}_{- 0.2}$ & 0.0 $^{+ 0.4}_{- 0.0}$ & 0.1 $^{+ 1.9}_{- 0.1}$ &
  0.0 $^{+ 0.0}_{- 0.5}$ & 0.1 $^{+ 2.0}_{- 0.1}$ & 4.0 $^{+ 4.5}_{-
  4.0}$ & $-0.1 ^{+ 0.1}_{- 0.4}$ & 4.0 $^{+ 4.5}_{- 4.0}$ \\
 
   $\Delta(2000)F_{35}$** & $5.3\pm 2.3$ & & & & & & & & \\
 
& & & $f_{\frac{1}{2}}$ & $f_{\frac{3}{2}}$ & $h_{\frac{3}{2}}$ & &
$f$ & $h$ \\ \cline{4-6} \cline{8-9}

$[\Delta\textstyle{7\over 2}^+]_1(1940)$ & 7.1 $\pm $ 0.1 & 1.2 $\pm
  $ 0.1 & 0.0 $\pm $ 0.0 & 0.0 $\pm $ 0.0 & 0.0 $\pm $ 0.0 & 0.0 $\pm
  $ 0.0 & -0.1 $\pm $ 0.0 & 0.0 $\pm $ 0.0 & 0.1 $\pm $ 0.0 \\
 
   $\Delta(1950)F_{37}$**** & 10.4 $\pm $ 1.1 & $\simeq$+1.5 $\pm $
   0.4 & & & & & & & \\
\tableline
\noalign{a\ \ Second $P_{13}$ found in Ref.~\cite{MANSA}.}
\noalign{b\ \ First $P_{31}$ state found in Ref.~\protect{\cite{MANSA}}.}
\noalign{c\ \ Ref.~\protect{\cite{MANSA}} finds two $F_{35}$ states;
this one and $\Delta(1905)F_{35}$.}
\end{tabular}
\end{table}
\begin{table}
\caption{Results for $N$ states in the $N=1$ and $N=2$ bands for
decays into the $\Lambda K$, $\Lambda K^*$, $\Lambda(1405) K$ and
$\Lambda(1520) K$ channels. Notation as in
Table~\protect{\ref{NDle2SK}}.}
\label{NDle2LK}
\begin{tabular}{lrrrrrrrrr}

 State &  $\Lambda K$ &  $\Lambda K^*$ &  $\Lambda K^*$ & 
  $\Lambda K^*$ &  $\sqrt{\Gamma_{\Lambda K^*}}$ & 
  $\Lambda(1405) K$ &  $\Lambda(1520) K$ &  $\Lambda(1520) K$ & 
  $\sqrt{\Gamma_{\Lambda(1520) K}}$ \\

 $[N\textstyle{1\over 2}^-]_2(1535)$ & $-5.2 ^{+ 1.4}_{- 0.5}$ \\
 
   $N(1650)S_{11}$**** & +3.3 $\pm $ 1.0 & & & & & & & & \\
 
 $[N\textstyle{3\over 2}^-]_1(1495)$ & 0.0 $^{+ 0.0}_{- 0.9}$ \\
 
   $N(1520)D_{13}$**** & 0.0 $\pm $ 0.0 & & & & & & & & \\
 
 $[N\textstyle{3\over 2}^-]_2(1625)$ & $-0.4 \pm $ 0.2 \\
 
   $N(1700)D_{13}$*** & $- 0.4 \pm $ 0.3 & & & & & & & & \\
 
 $[N\textstyle{5\over 2}^-]_1(1630)$ & 0.0 $\pm $ 0.0 \\
 
   $N(1675)D_{15}$**** & 0.4 $\pm $ 0.3 & & & & & & & & \\
 
 $[N\textstyle{1\over 2}^+]_3(1770)$ & $-2.8 \pm $ 0.6 \\
 
   $N(1710)P_{11}$*** & $+4.7 \pm $ 3.7 & & & & & & & & \\
 
& & $p_{\frac{1}{2}}$ & $p_{\frac{3}{2}}$ & & & & $d$\\ \cline{3-4}
\cline{8-8}

 $[N\textstyle{1\over 2}^+]_4(1880)$ & $-0.1 \pm $ 0.1 & 0.0 $\pm $
  0.0 & 0.0 $^{+ 0.2}_{- 0.0}$ & & 0.0 $^{+ 0.2}_{- 0.0}$ & 1.4 $^{+
  4.9}_{- 1.4}$ \\
 
 $[N\textstyle{1\over 2}^+]_5(1975)$ & $-1.1 ^{+ 0.3}_{- 0.2}$ & 0.1
  $^{+ 0.3}_{- 0.1}$ & 0.2 $^{+ 0.9}_{- 0.2}$ & & 0.2 $^{+ 0.9}_{-
  0.2}$ & $-0.1 \pm $ 0.1 & 0.0 $^{+ 0.0}_{- 0.2}$ & & 0.0 $^{+
  0.2}_{- 0.0}$ \\
 
& & $p_{\frac{1}{2}}$ & $p_{\frac{3}{2}}$ & $f_{\frac{3}{2}}$ & & &
$s$ & $d$\\ \cline{3-5} \cline{8-9}

 $[N\textstyle{3\over 2}^+]_1(1795)$ & $-4.3 ^{+ 0.8}_{- 0.7}$ \\
 
   $N(1720)P_{13}$**** & $+3.2 \pm $ 1.8 & & & & & & & & \\
 
 $[N\textstyle{3\over 2}^+]_2(1870)$ & $-0.9 ^{+ 0.4}_{- 0.1}$ & 0.0
  $^{+ 0.2}_{- 0.0}$ & 0.0 $^{+ 0.0}_{- 0.2}$ & 0.0 $\pm $ 0.0 & 0.0
  $^{+ 0.2}_{- 0.0}$ & 0.0 $^{+ 0.0}_{- 0.4}$ & 0.0 $^{+ 0.0}_{- 0.8}$
  & 0.0 $\pm $ 0.0 & 0.0 $^{+ 0.8}_{- 0.0}$ \\
 
 $[N\textstyle{3\over 2}^+]_3(1910)$ & 0.0 $\pm $ 0.0 & 0.0 $^{+
  0.1}_{- 0.0}$ & 0.0 $^{+ 0.0}_{- 0.2}$ & 0.0 $\pm $ 0.0 & 0.0 $^{+
  0.3}_{- 0.0}$ & 0.0 $^{+ 0.0}_{- 0.2}$ & 0.0 $\pm $ 0.1 & 0.0 $^{+
  0.0}_{- 0.2}$ & 0.0 $^{+ 0.2}_{- 0.0}$ \\
 
 $[N\textstyle{3\over 2}^+]_4(1950)$ & $-1.9 ^{+ 0.5}_{- 0.2}$ & 0.1
  $^{+ 0.5}_{- 0.1}$ & $-0.1 ^{+ 0.1}_{- 0.4}$ & 0.0 $^{+ 0.0}_{-
  0.1}$ & 0.1 $^{+ 0.6}_{- 0.1}$ & $-0.1 ^{+ 0.1}_{- 0.4}$ & 0.0 $^{+
  0.0}_{- 2.2}$ & 0.0 $^{+ 0.3}_{- 0.0}$ & 0.0 $^{+ 2.2}_{- 0.0}$ \\
 
 $[N\textstyle{3\over 2}^+]_5(2030)$ & $-0.9 \pm $ 0.2 & 0.1 $^{+
  0.3}_{- 0.1}$ & $-0.1 ^{+ 0.1}_{- 0.3}$ & 0.0 $^{+ 0.0}_{- 0.2}$ &
  0.1 $^{+ 0.4}_{- 0.1}$ & 0.1 $\pm $ 0.1 & $-0.6 ^{+ 0.6}_{- 0.3}$ &
  0.0 $\pm $ 0.0 & 0.6 $^{+ 0.3}_{- 0.6}$ \\
 
& & $f_{\frac{1}{2}}$ & $p_{\frac{3}{2}}$ & $f_{\frac{3}{2}}$ & & &
$d$ & $g$\\ \cline{3-5} \cline{8-9}

 $[N\textstyle{5\over 2}^+]_1(1770)$ & $-0.1 \pm $ 0.0 \\
 
   $N(1680)F_{15}$**** & $\simeq$ 0.1 $\pm $ 0.1 & & & & & & & & \\
 
 $[N\textstyle{5\over 2}^+]_2(1980)$ & 0.0 $\pm $ 0.0 & & & & 0.0 $\pm
  $ 0.0 & -0.3 $^{+ 0.2}_{- 0.4}$ & & & 0.0 $\pm $ 0.0 \\
 
 $[N\textstyle{5\over 2}^+]_3(1995)$ & $-0.5 \pm $ 0.3 & 0.0 $\pm $
  0.1 & 0.3 $^{+ 1.2}_{- 0.2}$ & 0.0 $^{+ 0.2}_{- 0.0}$ & 0.3 $^{+
  1.3}_{- 0.2}$ & $-0.6 ^{+ 0.6}_{- 1.6}$ & 0.0 $^{+ 0.0}_{- 0.5}$ &
  0.0 $\pm $ 0.0 & 0.0 $^{+ 0.5}_{- 0.0}$ \\
 
& & & & & & & $d$ & $g$\\ \cline{8-9}

 $[N\textstyle{7\over 2}^+]_1(2000)$ & 0.0 $\pm $ 0.0 & & & & 0.0 $\pm
  $ 0.0 & 0.0 $\pm $ 0.0 & 0.0 $^{+ 0.0}_{- 0.5}$ & 0.0 $\pm $ 0.0 &
  0.0 $^{+ 0.5}_{- 0.0}$ \\
 
   & $\simeq$ 1.5 $\pm $ 2.4 & & & & & & & & \\

\end{tabular}
\end{table}
\begin{table}
\caption{Results in the $\Sigma K$, $\Sigma K^*$, and $\Sigma(1385) K$
channels for the lightest few negative-parity $N$ and $\Delta$
resonances of each $J$ in the N=3 band, and for the lightest few $N$
and $\Delta$ resonances for $J^P$ values which first appear in the
N=4, 5 and 6 bands. Notation as in Table~\protect{\ref{NDle2SK}}.}
\label{NDge3SK}
\begin{tabular}{lrrrrrrrrr}
 State & $N \pi$ & $\Sigma K$ & $\Sigma K^*$ & $\Sigma K^*$ & $\Sigma
  K^*$ & $\sqrt{\Gamma_{\Sigma K^*}}$ & $\Sigma(1385) K$ &
  $\Sigma(1385) K$ & $\sqrt{\Gamma_{\Sigma(1385) K}}$\\

& & & $s_{\frac{1}{2}}$ & $d_{\frac{1}{2}}$ & & & $d$ & \\
\cline{4-5} \cline{8-8}

 $[N\textstyle{1\over 2}^-]_3(1945)$ & 5.7 $_{- 1.6}^{+ 0.5}$ & 2.1
  $^{+ 1.3}_{- 1.4}$ & $-0.9 ^{+ 0.8}_{- 0.7}$ & $-0.2 ^{+ 0.2}_{-
  1.3}$ & & 0.9 $^{+ 1.2}_{- 0.8}$ & 1.7 $^{+ 2.0}_{- 1.4}$ & & 1.7
  $^{+ 2.0}_{- 1.4}$ \\
 
   $N(2090)S_{11}$* & $7.9\pm 3.8$ \\
 
 $[N\textstyle{1\over 2}^-]_4(2030)$ & $3.7 _{- 1.1}^{+ 0.5}$ & $-4.5
  ^{+ 2.8}_{- 2.4}$ & $-0.7 ^{+ 0.6}_{- 2.7}$ & 0.1 $^{+ 0.6}_{-
  0.1}$ & & 0.7 $^{+ 2.8}_{- 0.6}$ & 1.0 $^{+ 1.0}_{- 0.9}$ & & 1.0
  $^{+ 1.0}_{- 0.9}$ \\
 
 $[N\textstyle{1\over 2}^-]_5(2070)$ & $2.1 _{- 1.5}^{+ 0.8}$ & $-1.5
  \pm $ 0.6 & 2.9 $^{+ 5.7}_{- 2.6}$ & 0.1 $^{+ 1.1}_{- 0.1}$ & & 2.9
  $^{+ 5.7}_{- 2.6}$ & $-3.3 ^{+ 2.9}_{- 2.7}$ & & 3.3 $^{+ 2.7}_{-
  2.9}$ \\
 
 $[N\textstyle{1\over 2}^-]_6(2145)$ & 0.4 $\pm $ 0.0 & $-1.1 \pm $ 0.7
  & 0.0 $\pm $ 0.0 & 0.3 $^{+ 0.8}_{- 0.3}$ & & 0.3 $^{+ 0.8}_{- 0.3}$
  & $-0.2 ^{+ 0.1}_{- 0.5}$ & & 0.2 $^{+ 0.5}_{- 0.1}$ \\
 
 $[N\textstyle{1\over 2}^-]_7(2195)$ & $0.1 \pm $ 0.1 & $-0.7 \pm $ 0.9
  & 0.5 $\pm $ 0.3 & 0.5 $^{+ 0.8}_{- 0.4}$ & & 0.7 $^{+ 0.8}_{- 0.5}$
  & $-0.8 ^{+ 0.3}_{- 0.9}$ & & 0.8 $^{+ 0.9}_{- 0.3}$ \\
 
 $[N\textstyle{3\over 2}^-]_3(1960)$ & 8.2 $^{+ 0.7}_{- 1.7}$ & $-0.7
  \pm $ 0.3 & 0.1 $^{+ 0.9}_{- 0.1}$ & $-0.5 \pm $ 0.5 & 0.0 $^{+
  0.0}_{- 0.3}$ & 0.6 $^{+ 1.0}_{- 0.5}$ & 1.3 $\pm $ 0.4 & 1.4 $\pm $
  1.3 & 1.9 $^{+ 1.3}_{- 1.0}$ \\
 
   $N(2080)D_{13}$** & $6.2 \pm $ 2.0 & $\simeq$ 1.2 $\pm $ 0.0 & & &
   & & & & \\
 
& & & $d_{\frac{1}{2}}$ & $s_{\frac{3}{2}}$ & $d_{\frac{3}{2}}$ & &
$s$ & $d$ \\ \cline{4-6} \cline{8-9}

 $[N\textstyle{3\over 2}^-]_4(2055)$ & 6.2 $_{- 0.6}^{+ 0.1}$ & 1.8
  $^{+ 0.7}_{- 0.8}$ & $-0.2 ^{+ 0.2}_{- 1.3}$ & 1.2 $^{+ 2.9}_{-
  1.1}$ & 0.0 $\pm $ 0.1 & 1.2 $^{+ 3.2}_{- 1.1}$ & $-2.5 \pm $ 1.0 &
  $-2.5 ^{+ 2.3}_{- 1.9}$ & 3.5 $\pm $ 2.1 \\
 
 $[N\textstyle{3\over 2}^-]_5(2095)$ & 0.2 $\pm $ 0.2 & 0.4 $\pm $ 0.1
  & 0.3 $^{+ 1.9}_{- 0.3}$ & 1.7 $^{+ 1.2}_{- 1.5}$ & $-0.4 ^{+
  0.3}_{- 2.3}$ & 1.7 $^{+ 2.8}_{- 1.6}$ & 7.7 $\pm $ 1.2 & $-0.8 ^{+
  0.7}_{- 1.0}$ & 7.8 $^{+ 1.3}_{- 1.2}$ \\
 
 $[N\textstyle{3\over 2}^-]_6(2165)$ & 1.5 $_{- 0.2}^{+ 0.1}$ & 2.4
  $^{+ 0.5}_{- 0.7}$ & 0.2 $^{+ 0.4}_{- 0.2}$ & $-0.1 \pm $ 0.0 & 0.2
  $^{+ 0.4}_{- 0.2}$ & 0.3 $^{+ 0.6}_{- 0.3}$ & 0.0 $\pm $ 0.1 & 0.4
  $^{+ 0.8}_{- 0.3}$ & 0.4 $^{+ 0.8}_{- 0.3}$ \\
 
 $[N\textstyle{3\over 2}^-]_7(2180)$ & 1.7 $_{- 0.2}^{+ 0.1}$ & 1.8
  $^{+ 0.3}_{- 0.5}$ & 0.4 $^{+ 0.6}_{- 0.3}$ & 0.0 $\pm $ 0.0 & 0.4
  $^{+ 0.7}_{- 0.4}$ & 0.5 $^{+ 1.0}_{- 0.5}$ & $-0.1 \pm $ 0.0 & 1.0
  $^{+ 1.6}_{- 0.5}$ & 1.0 $^{+ 1.6}_{- 0.5}$ \\
 
& & & $d_{\frac{1}{2}}$ & $d_{\frac{3}{2}}$ & $g_{\frac{3}{2}}$ & &
$d$ & $g$ \\ \cline{4-6} \cline{8-9}

 $[N\textstyle{5\over 2}^-]_2(2080)$ & $5.1 _{- 0.8}^{+ 0.2}$ & $-2.4
  ^{+ 0.9}_{- 0.5}$ & $-0.1 ^{+ 0.1}_{- 0.2}$ & 0.8 $^{+ 2.3}_{-
  0.8}$ & 0.0 $\pm $ 0.1 & 0.8 $^{+ 2.3}_{- 0.8}$ & 1.7 $^{+ 1.5}_{-
  1.2}$ & $-0.1 ^{+ 0.1}_{- 0.3}$ & 1.7 $^{+ 1.5}_{- 1.2}$ \\
 
 $[N\textstyle{5\over 2}^-]_3(2095)$ & 5.2 $^{+ 0.4}_{- 1.0}$ & 2.5 $^{+
  0.6}_{- 0.9}$ & 0.2 $^{+ 1.2}_{- 0.2}$ & $-0.2 ^{+ 0.2}_{- 1.5}$ &
  0.0 $\pm $ 0.0 & 0.3 $^{+ 1.9}_{- 0.3}$ & $-2.0 ^{+ 1.6}_{- 2.5}$ &
  0.0 $^{+ 0.0}_{- 0.1}$ & 2.0 $^{+ 2.5}_{- 1.6}$ \\
 
   $N(2200)D_{15}$** & 4.7 $\pm $ 1.0 \\
 
 $[N\textstyle{5\over 2}^-]_4(2180)$ & 1.9 $^{+ 0.1}_{- 0.3}$ & 1.5 $^{+
  0.3}_{- 0.4}$ & 0.2 $^{+ 0.3}_{- 0.2}$ & 0.5 $^{+ 0.8}_{- 0.4}$ &
  0.0 $^{+ 0.0}_{- 0.1}$ & 0.5 $^{+ 0.9}_{- 0.4}$ & 1.1 $^{+ 1.7}_{-
  0.5}$ & $-0.1 ^{+ 0.1}_{- 0.4}$ & 1.1 $^{+ 1.7}_{- 0.5}$ \\
 
 $[N\textstyle{5\over 2}^-]_5(2235)$ & 2.0 $^{+ 0.1}_{- 0.3}$ & 0.4 $\pm
  $ 0.1 & $-1.1 ^{+ 1.0}_{- 1.2}$ & $-1.0 ^{+ 0.9}_{- 1.1}$ & 0.0 $\pm
  $ 0.0 & 1.5 $^{+ 1.7}_{- 1.3}$ & $-0.6 \pm $ 0.3 & 0.6 $^{+ 0.7}_{-
  0.5}$ & 0.9 $^{+ 0.7}_{- 0.5}$ \\
 
 $[N\textstyle{5\over 2}^-]_6(2260)$ & 0.4 $\pm $ 0.1 & $-0.2 \pm $ 0.1
  & $-1.1 ^{+ 1.0}_{- 1.3}$ & 1.0 $^{+ 1.1}_{- 0.8}$ & $-0.1 ^{+
  0.1}_{- 0.3}$ & 1.5 $^{+ 1.7}_{- 1.3}$ & 2.8 $^{+ 0.2}_{- 1.7}$ &
  0.3 $\pm $ 0.2 & 2.8 $^{+ 0.2}_{- 1.7}$ \\
 
 $[N\textstyle{5\over 2}^-]_7(2295)$ & 0.2 $\pm $ 0.0 & 1.8 $^{+ 0.2}_{-
  0.3}$ & 0.2 $^{+ 0.3}_{- 0.1}$ & 0.4 $^{+ 0.6}_{- 0.3}$ & $-0.1 ^{+
  0.1}_{- 0.4}$ & 0.4 $^{+ 0.7}_{- 0.3}$ & 1.0 $\pm $ 0.5 & $-0.1 \pm
  $ 0.1 & 1.1 $\pm $ 0.5 \\
 
 $[N\textstyle{5\over 2}^-]_8(2305)$ & 0.3 $\pm $ 0.1 & 0.7 $\pm $ 0.0
  & $-0.7 ^{+ 0.5}_{- 1.2}$ & 0.3 $^{+ 0.4}_{- 0.2}$ & $-0.1 ^{+
  0.1}_{- 0.4}$ & 0.8 $^{+ 1.3}_{- 0.6}$ & 1.2 $\pm $ 0.6 & 0.4 $^{+
  0.1}_{- 0.3}$ & 1.3 $\pm $ 0.6 \\
 
& & & $g_{\frac{1}{2}}$ & $d_{\frac{3}{2}}$ & $g_{\frac{3}{2}}$ & &
$d$ & $g$ \\ \cline{4-6} \cline{8-9}

 $[N\textstyle{7\over 2}^-]_1(2090)$ & 6.9 $\pm $ 1.3 & $-0.2 \pm $ 0.1
  & 0.0 $\pm $ 0.1 & $-0.3 ^{+ 0.2}_{- 0.4}$ & 0.0 $\pm $ 0.0 & 0.3
  $^{+ 0.4}_{- 0.2}$ & 0.3 $^{+ 0.3}_{- 0.1}$ & 0.2 $^{+ 0.4}_{- 0.1}$
  & 0.3 $^{+ 0.5}_{- 0.1}$ \\
 
   $N(2190)G_{17}$**** & $7.0\pm 3.0$\\

 $[N\textstyle{7\over 2}^-]_2(2305)$ & 0.4 $\pm $ 0.1 & 0.4 $\pm $ 0.2 &
  $-0.1 ^{+ 0.1}_{- 0.2}$ & 0.8 $^{+ 1.2}_{- 0.6}$ & $-0.1 ^{+ 0.1}_{-
  0.2}$ & 0.8 $^{+ 1.3}_{- 0.6}$ & 0.0 $\pm $ 0.0 & $-0.6 ^{+ 0.5}_{-
  0.1}$ & 0.6 $^{+ 0.1}_{- 0.5}$ \\
 
 $[N\textstyle{7\over 2}^-]_3(2355)$ & 1.1 $\pm $ 0.3 & 0.0 $\pm $ 0.0 &
  0.0 $\pm $ 0.0 & 0.0 $\pm $ 0.0 & 0.0 $\pm $ 0.1 & 0.1 $\pm $ 0.1 &
  $-0.3 \pm $ 0.1 & 0.0 $\pm $ 0.0 & 0.3 $\pm $ 0.1 \\
 
& & & $g_{\frac{1}{2}}$ & $g_{\frac{3}{2}}$ & $i_{\frac{3}{2}}$ & &
$g$ & $i$ \\ \cline{4-6} \cline{8-9}

 $[N\textstyle{9\over 2}^-]_1(2215)$ & 2.5 $\pm $ 0.4 & 1.1 $\pm $ 0.4 &
  0.1 $\pm $ 0.1 & $-0.2 \pm $ 0.1 & 0.0 $\pm $ 0.0 & 0.2 $\pm $ 0.2 &
  $-0.7 \pm $ 0.4 & 0.0 $\pm $ 0.0 & 0.7 $\pm $ 0.5 \\
 
   $N(2250)G_{19}$**** & 6.1 $\pm $ 1.0 & & & &
   & & & & \\
 
& & & $i_{\frac{1}{2}}$ & $g_{\frac{3}{2}}$ & $i_{\frac{3}{2}}$ & &
$g$ & $i$ \\ \cline{4-6} \cline{8-9}

 $[N\textstyle{11\over 2}^-]_1(2600)$ & 3.3 $^{+ 1.1}_{- 0.9}$ & 0.0
  $\pm $ 0.0 & 0.0 $\pm $ 0.0 & $-0.1 \pm $ 0.0 & 0.0 $\pm $ 0.0 & 0.1
  $\pm $ 0.1 & 0.0 $\pm $ 0.0 & 0.0 $\pm $ 0.1 & 0.1 $\pm $ 0.1 \\
 
   $N(2600)I_{1\,\,11}$*** & $4.5\pm 1.5$\\
 
 $[N\textstyle{11\over 2}^-]_2(2670)$ & 1.8 $\pm $ 0.5 & 0.2 $\pm $ 0.1
  & 0.0 $\pm $ 0.0 & 0.5 $^{+ 0.7}_{- 0.2}$ & $-0.1 ^{+ 0.1}_{- 0.2}$
  & 0.5 $^{+ 0.7}_{- 0.2}$ & 0.1 $\pm $ 0.1 & $-0.2 ^{+ 0.1}_{- 0.5}$
  & 0.3 $^{+ 0.5}_{- 0.1}$ \\
 
 $[N\textstyle{11\over 2}^-]_3(2700)$ & 0.3 $\pm $ 0.1 & 0.0 $\pm $ 0.0
  & $-0.2 ^{+ 0.1}_{- 0.5}$ & 0.0 $\pm $ 0.0 & 0.2 $^{+ 0.4}_{- 0.1}$
  & 0.3 $^{+ 0.6}_{- 0.2}$ & $-0.9 ^{+ 0.4}_{- 1.0}$ & 0.0 $\pm $ 0.0
  & 0.9 $^{+ 1.0}_{- 0.4}$ \\
 
 $[N\textstyle{11\over 2}^-]_4(2770)$ & 0.2 $\pm $ 0.0 & 0.2 $\pm $ 0.1
  & $-0.1 \pm $ 0.1 & 0.5 $^{+ 0.4}_{- 0.3}$ & $-0.1 \pm $ 0.1 & 0.5
  $^{+ 0.5}_{- 0.3}$ & $-0.1 \pm $ 0.1 & $-0.3 ^{+ 0.2}_{- 0.5}$ & 0.3
  $^{+ 0.5}_{- 0.2}$ \\
 
 $[N\textstyle{11\over 2}^-]_5(2855)$ & 0.6 $\pm $ 0.1 & 0.0 $\pm $ 0.0
  & $-0.1 \pm $ 0.1 & 0.1 $\pm $ 0.0 & 0.0 $\pm $ 0.0 & 0.1 $\pm $ 0.1
  & $-0.2 \pm $ 0.1 & $-0.1 \pm $ 0.0 & 0.2 $\pm $ 0.1 \\
 
& & & $i_{\frac{1}{2}}$ & $i_{\frac{3}{2}}$ & $k_{\frac{3}{2}}$ & &
$i$ & $k$ \\ \cline{4-6} \cline{8-9}

 $[N\textstyle{13\over 2}^-]_1(2715)$ & 1.1 $\pm $ 0.3 & 0.4 $\pm $ 0.2
  & 0.1 $^{+ 0.3}_{- 0.1}$ & $-0.2 ^{+ 0.1}_{- 0.4}$ & 0.0 $\pm $ 0.0
  & 0.2 $^{+ 0.5}_{- 0.2}$ & $-0.3 ^{+ 0.2}_{- 0.6}$ & 0.0 $\pm $ 0.0
  & 0.3 $^{+ 0.6}_{- 0.2}$ \\
 
 $[N\textstyle{13\over 2}^-]_2(2845)$ & 0.2 $\pm $ 0.1 & 0.1 $\pm $
  0.1 & 0.1 $\pm $ 0.1 & $-0.1 \pm $ 0.1 & 0.0 $\pm $ 0.0 & 0.2 $\pm $
  0.1 & $-0.2 ^{+ 0.1}_{- 0.2}$ & 0.0 $\pm $ 0.0 & 0.2 $^{+ 0.2}_{-
  0.1}$ \\
 
& & & $s_{\frac{1}{2}}$ & $d_{\frac{1}{2}}$ & & & $d$ & \\
\cline{4-5} \cline{8-8}

$[\Delta\textstyle{1\over 2}^-]_2(2035)$ & 1.2 $\pm $ 0.2 & -1.9 $\pm $
  0.3 & 0.1 $\pm $ 0.2 & 0.0 $\pm $ 0.0 & & 0.1 $\pm $ 0.2 & $-0.3 ^{+
  0.2}_{- 0.6}$ & & 0.3 $^{+ 0.6}_{- 0.2}$ \\
 
   $\Delta(1900)S_{31}$*** & 6.3 $\pm $ 1.0 & $<$ 0.9 & & & &
   & & & \\
 
$[\Delta\textstyle{1\over 2}^-]_3(2140)$ & 3.1 $^{+ 0.4}_{- 1.1}$ & $-4.1
  \pm $ 2.4 & $-4.8 ^{+ 4.2}_{- 0.7}$ & 0.4 $^{+ 1.0}_{- 0.4}$ & &
  4.8 $^{+ 0.9}_{- 4.2}$ & 1.4 $^{+ 2.9}_{- 0.9}$ & & 1.4 $^{+ 2.9}_{-
  0.9}$ \\
 
   $\Delta(2150)S_{31}$* & 6.6 $\pm $ 2.0 & $<$ 0.7 & & & & & & & \\
 
& & & $d_{\frac{1}{2}}$ & $s_{\frac{3}{2}}$ & $d_{\frac{3}{2}}$ & &
$s$ & $d$ \\ \cline{4-6} \cline{8-9}

$[\Delta\textstyle{3\over 2}^-]_2(2080)$ & 2.1 $\pm $ 0.1 & 1.1 $\pm $
  0.7 & 0.0 $^{+ 0.0}_{- 0.4}$ & 0.1 $^{+ 1.0}_{- 0.1}$ & 0.0 $^{+
  0.1}_{- 0.0}$ & 0.1 $^{+ 1.1}_{- 0.1}$ & $-4.1 ^{+ 4.0}_{- 1.5}$ &
  $-0.5 ^{+ 0.5}_{- 2.2}$ & 4.2 $^{+ 2.1}_{- 4.0}$ \\
 
   $\Delta(1940)D_{33}$* & 6.5 $\pm $ 2.0 & $<$ 0.7 & & & & & & & \\
 
$[\Delta\textstyle{3\over 2}^-]_3(2145)$ & 2.2 $^{+ 0.1}_{- 0.3}$ &
  1.9 $^{+ 0.5}_{- 0.6}$ & 0.1 $\pm $ 0.1 & 5.7 $^{+ 1.2}_{- 5.0}$ &
  $-0.6 ^{+ 0.6}_{- 1.7}$ & 5.7 $^{+ 1.5}_{- 5.0}$ & 5.2 $\pm $ 0.4 &
  $-1.9 ^{+ 1.2}_{- 4.0}$ & 5.5 $^{+ 2.6}_{- 0.7}$ \\
 
& & & $d_{\frac{1}{2}}$ & $d_{\frac{3}{2}}$ & $g_{\frac{3}{2}}$ & &
$d$ & $g$ \\ \cline{4-6} \cline{8-9}

$[\Delta\textstyle{5\over 2}^-]_1(2155)$ & 5.2 $\pm $ 0.0 & 2.1 $\pm $
  0.4 & & & & 0.0 $\pm $ 0.0 & $-0.2 ^{+ 0.2}_{- 0.3}$ & 0.0 $\pm $
  0.0 & 0.2 $^{+ 0.3}_{- 0.2}$ \\
 
   $\Delta(1930)D_{35}$*** & 7.2 $\pm $ 1.0 & $<$ 0.7 & & &
   & & & & \\
 
$[\Delta\textstyle{5\over 2}^-]_2(2165)$ & 0.6 $\pm $ 0.1 & 1.0 $\pm $
  0.3 & 0.9 $^{+ 1.8}_{- 0.9}$ & $-0.4 ^{+ 0.3}_{- 0.8}$ & 0.0 $^{+
  0.1}_{- 0.0}$ & 1.0 $^{+ 2.0}_{- 0.9}$ & $-1.9 ^{+ 1.1}_{- 3.7}$ &
  $-0.3 ^{+ 0.2}_{- 1.0}$ & 2.0 $^{+ 3.8}_{- 1.1}$ \\
 
$[\Delta\textstyle{5\over 2}^-]_3(2265)$ & 2.4 $\pm $ 0.4 & 2.5 $\pm $
  0.1 & 0.8 $^{+ 0.9}_{- 0.2}$ & 1.0 $^{+ 1.1}_{- 0.3}$ & $-0.2 ^{+
  0.1}_{- 0.4}$ & 1.3 $^{+ 1.4}_{- 0.3}$ & 1.9 $\pm $ 0.3 & $-0.6 ^{+
  0.2}_{- 0.1}$ & 2.0 $\pm $ 0.3 \\
 
   $\Delta(2350)D_{35}$* & 5.1 $\pm $ 2.0 &
   $<$ 0.9 & & & & & & & \\
 
$[\Delta\textstyle{5\over 2}^-]_4(2325)$ & 0.1 $\pm $ 0.0 & $-0.3 \pm $
  0.1 & 0.3 $^{+ 0.4}_{- 0.2}$ & 1.5 $^{+ 2.4}_{- 1.0}$ & $-0.1 ^{+
  0.1}_{- 0.2}$ & 1.6 $^{+ 2.4}_{- 1.0}$ & 1.8 $\pm $ 0.8 & $-0.5 ^{+
  0.4}_{- 0.1}$ & 1.9 $\pm $ 0.8 \\
 
& & & $g_{\frac{1}{2}}$ & $d_{\frac{3}{2}}$ & $g_{\frac{3}{2}}$ & &
$d$ & $g$ \\ \cline{4-6} \cline{8-9}

$[\Delta\textstyle{7\over 2}^-]_1(2230)$ & 2.1 $\pm $ 0.6 & 0.4 $^{+
  0.3}_{- 0.2}$ & $-0.1 ^{+ 0.1}_{- 0.4}$ & 0.5 $^{+ 0.8}_{- 0.5}$ &
  0.0 $\pm $ 0.1 & 0.5 $^{+ 0.9}_{- 0.5}$ & $-2.0 ^{+ 0.8}_{- 2.1}$ &
  $-0.3 ^{+ 0.2}_{- 0.8}$ & 2.0 $^{+ 2.2}_{- 0.9}$ \\
 
   $\Delta(2200)G_{37}$* & 5.4 $\pm $ 1.0 &
   $\simeq$- 1.1 $\pm $ 0.0 & & & & & & & \\
 
$[\Delta\textstyle{7\over 2}^-]-2(2295)$ & 1.8 $\pm $ 0.4 & 0.5 $\pm $
  0.3 & 0.1 $^{+ 0.3}_{- 0.1}$ & 1.3 $^{+ 2.0}_{- 1.0}$ & $-0.2 ^{+
  0.2}_{- 0.8}$ & 1.4 $^{+ 2.1}_{- 1.0}$ & 2.5 $\pm $ 1.3 & $-0.7 ^{+
  0.6}_{- 0.2}$ & 2.6 $\pm $ 1.3 \\
 
& & & $g_{\frac{1}{2}}$ & $g_{\frac{3}{2}}$ & $i_{\frac{3}{2}}$ & &
$g$ & $i$ \\ \cline{4-6} \cline{8-9}

$[\Delta\textstyle{9\over 2}^-]_1(2295)$ & 4.8 $\pm $ 1.3 & 1.4 $^{+
  1.0}_{- 0.8}$ & 0.4 $^{+ 0.5}_{- 0.3}$ & $-0.6 ^{+ 0.5}_{- 0.8}$ &
  0.0 $\pm $ 0.0 & 0.7 $^{+ 0.9}_{- 0.6}$ & $-0.9 \pm $ 0.7 & 0.0 $\pm
  $ 0.0 & 0.9 $\pm $ 0.7 \\
 
   $\Delta(2400)G_{39}$** & 5.4 $\pm $ 1.0 &
   $<$ 1.2 & & & & & & & \\
 
& & & $i_{\frac{1}{2}}$ & $i_{\frac{3}{2}}$ & $k_{\frac{3}{2}}$ & &
$i$ & $k$ \\ \cline{4-6} \cline{8-9}

$[\Delta\textstyle{13\over 2}^-]_1(2750)$ & 2.2 $\pm $ 0.4 & 0.4 $\pm $
  0.1 & 0.2 $^{+ 0.2}_{- 0.1}$ & $-0.3 ^{+ 0.1}_{- 0.3}$ & 0.0 $\pm $
  0.0 & 0.3 $^{+ 0.4}_{- 0.2}$ & $-0.4 ^{+ 0.2}_{- 0.4}$ & 0.0 $\pm $
  0.0 & 0.4 $^{+ 0.4}_{- 0.2}$ \\
 
   $\Delta(2750)I_{3\,\,13}$** & $3.7\pm 1.5$\\
 
& & & $p_{\frac{1}{2}}$ & $p_{\frac{3}{2}}$ & & & $p$ \\ \cline{4-5}
\cline{8-8}

 $[N\textstyle{1\over 2}^+]_6(2065)$ & 7.7 $_{- 2.9}^{+ 2.4}$ & 0.3
  $\pm $ 0.3 & 0.1 $\pm $ 0.0 & $-0.1 \pm $ 0.0 & & 0.1 $\pm $ 0.0 &
  0.6 $\pm $ 0.2 & & 0.6 $\pm $ 0.2 \\
 
   $N(2100)P_{11}$* & $5.0\pm 2.0$\\
 
 $[N\textstyle{1\over 2}^+]_7(2210)$ & 0.3 $_{- 0.1}^{+ 0.2}$ & 0.0
  $^{+ 0.7}_{- 0.5}$ & 0.5 $^{+ 0.1}_{- 0.3}$ & $-0.3 \pm $ 0.2 & &
  0.5 $^{+ 0.2}_{- 0.4}$ & 1.1 $^{+ 0.6}_{- 0.4}$ & & 1.1 $^{+ 0.6}_{-
  0.4}$ \\
 
& & & $f_{\frac{1}{2}}$ & $f_{\frac{3}{2}}$ & $h_{\frac{3}{2}}$ & &
$f$ & $h$ \\ \cline{4-6} \cline{8-9}

 $[N\textstyle{7\over 2}^+]_2(2390)$ & 4.9 $^{+ 0.0}_{- 0.4}$ & 0.1 $\pm
  $ 0.0 & $-0.2 \pm $ 0.1 & $-0.2 \pm $ 0.1 & 0.0 $\pm $ 0.0 & 0.3
  $\pm $ 0.2 & 0.0 $\pm $ 0.0 & 0.1 $\pm $ 0.1 & 0.1 $\pm $ 0.1 \\
 
 $[N\textstyle{7\over 2}^+]_3(2410)$ & 0.4 $^{+ 0.2}_{- 0.4}$ & 1.7 $^{+
  0.3}_{- 0.4}$ & 0.7 $^{+ 0.3}_{- 0.4}$ & $-1.2 ^{+ 0.8}_{- 0.5}$ &
  0.0 $\pm $ 0.0 & 1.4 $^{+ 0.6}_{- 0.9}$ & $-1.6 \pm $ 0.3 & 0.0 $\pm
  $ 0.0 & 1.6 $\pm $ 0.3 \\
 
 $[N\textstyle{7\over 2}^+]_4(2455)$ & 0.5 $\pm $ 0.0 & 0.5 $\pm $ 0.1 &
  $-1.1 ^{+ 0.8}_{- 0.1}$ & 0.3 $^{+ 0.0}_{- 0.2}$ & $-0.2 \pm $ 0.1 &
  1.1 $^{+ 0.1}_{- 0.8}$ & 1.1 $\pm $ 0.1 & 0.2 $^{+ 0.2}_{- 0.1}$ &
  1.1 $\pm $ 0.2 \\
 
& & & $h_{\frac{1}{2}}$ & $f_{\frac{3}{2}}$ & $h_{\frac{3}{2}}$ & &
$f$ & $h$ \\ \cline{4-6} \cline{8-9}

 $[N\textstyle{9\over 2}^+]_1(2345)$ & 3.6 $^{+ 1.0}_{- 0.8}$ & 0.0
  $\pm $ 0.0 & & & & 0.0 $\pm $ 0.0 & $-0.1 \pm $ 0.0 & 0.0 $\pm $ 0.0
  & 0.1 $\pm $ 0.1 \\
 
   $N(2220)H_{19}$**** & 8.1 $\pm $ 1.0 & & & &
   & & & & \\
 
 $[N\textstyle{9\over 2}^+]_2(2500)$ & 0.4 $\pm $ 0.1 & $-0.4 \pm $ 0.2
  & 0.2 $\pm $ 0.1 & $-1.3 ^{+ 0.8}_{- 0.2}$ & 0.1 $\pm $ 0.1 & 1.3
  $^{+ 0.3}_{- 0.8}$ & 0.2 $\pm $ 0.0 & 0.4 $^{+ 0.5}_{- 0.1}$ & 0.5
  $^{+ 0.4}_{- 0.1}$ \\
 
 $[N\textstyle{9\over 2}^+]_3(2490)$ & 0.6 $\pm $ 0.2 & 0.0 $\pm $ 0.0 &
  $-0.3 \pm $ 0.2 & -0.1 $\pm $ 0.1 & 0.2 $\pm $ 0.2 & 0.4 $\pm $ 0.3
  & $-1.7 \pm $ 0.3 & 0.0 $\pm $ 0.0 & 1.7 $\pm $ 0.3 \\
 
& & & $h_{\frac{1}{2}}$ & $h_{\frac{3}{2}}$ & $j_{\frac{3}{2}}$ & &
$h$ & $j$ \\ \cline{4-6} \cline{8-9}

 $[N\textstyle{11\over 2}^+]_1(2490)$ & 1.3 $\pm $ 0.4 & 0.6 $\pm $ 0.4
  & 0.1 $\pm $ 0.1 & $-0.2 \pm $ 0.2 & 0.0 $\pm $ 0.0 & 0.3 $\pm $ 0.2
  & $-0.3 ^{+ 0.1}_{- 0.3}$ & 0.0 $\pm $ 0.0 & 0.3 $^{+ 0.3}_{- 0.1}$
  \\
 
 $[N\textstyle{11\over 2}^+]_2(2600)$ & 0.7 $\pm $ 0.1 & 0.3 $\pm $ 0.1
  & 0.1 $\pm $ 0.1 & $-0.1 ^{+ 0.0}_{- 0.2}$ & 0.0 $\pm $ 0.0 & 0.1
  $^{+ 0.2}_{- 0.1}$ & $-0.2 ^{+ 0.1}_{- 0.2}$ & 0.0 $\pm $ 0.0 & 0.2
  $^{+ 0.2}_{- 0.1}$ \\
 
& & & $j_{\frac{1}{2}}$ & $h_{\frac{3}{2}}$ & $j_{\frac{3}{2}}$ & &
$h$ & $j$ \\ \cline{4-6} \cline{8-9}

 $[N\textstyle{13\over 2}^+]_1(2820)$ & 2.0 $_{- 0.6}^{+ 0.8}$ & 0.0
  $\pm $ 0.0 & & & & 0.0 $\pm $ 0.0 & & & 0.0 $\pm $ 0.0 \\
 
   $N(2700)K_{1\,\,13}$** & $3.7\pm 1.2$\\
 
 $[N\textstyle{13\over 2}^+]_2(2930)$ & 0.2 $\pm $ 0.1 & $-0.1 \pm $
  0.1 & 0.1 $\pm $ 0.0 & $-0.6 ^{+ 0.3}_{- 0.1}$ & 0.1 $\pm $ 0.1 &
  0.7 $^{+ 0.1}_{- 0.3}$ & 0.0 $\pm $ 0.0 & 0.5 $\pm $ 0.4 & 0.5 $\pm
  $ 0.4 \\
 
 $[N\textstyle{13\over 2}^+]_3(2955)$ & 0.2 $\pm $ 0.1 & 0.0 $\pm $ 0.0
  & $-0.3 \pm $ 0.2 & 0.1 $\pm $ 0.0 & 0.2 $\pm $ 0.1 & 0.4 $\pm $ 0.2
  & $-1.2 ^{+ 0.6}_{- 0.4}$ & $-0.1 \pm $ 0.1 & 1.2 $^{+ 0.4}_{- 0.6}$
  \\
 
& & & $j_{\frac{1}{2}}$ & $j_{\frac{3}{2}}$ & $l_{\frac{3}{2}}$ & &
$j$ & $l$ \\ \cline{4-6} \cline{8-9}

 $[N\textstyle{15\over 2}^+]_1(2940)$ & 0.7 $\pm $ 0.2 & 0.2 $\pm $ 0.1
  & 0.1 $\pm $ 0.1 & $-0.2 \pm $ 0.1 & 0.0 $\pm $ 0.0 & 0.3 $\pm $ 0.1
  & $-0.4 \pm $ 0.3 & 0.0 $\pm $ 0.0 & 0.4 $\pm $ 0.3 \\
 
 $[N\textstyle{15\over 2}^+]_2(3005)$ & 0.4 $\pm $ 0.1 & 0.1 $\pm $ 0.1
  & 0.1 $\pm $ 0.0 & $-0.1 \pm $ 0.0 & 0.0 $\pm $ 0.0 & 0.1 $\pm $ 0.0
  & $-0.2 \pm $ 0.1 & 0.0 $\pm $ 0.0 & 0.2 $\pm $ 0.1 \\
 
& & & $f_{\frac{1}{2}}$ & $f_{\frac{3}{2}}$ & $h_{\frac{3}{2}}$ & &
$f$ & $h$ \\ \cline{4-6} \cline{8-9}

$[\Delta\textstyle{7\over 2}^+]_2(2370)$ & 1.5 $^{+ 0.6}_{- 0.9}$ & 1.9
  $^{+ 0.4}_{- 0.5}$ & 0.6 $^{+ 0.6}_{- 0.4}$ & $-1.0 ^{+ 0.7}_{-
  1.1}$ & 0.0 $\pm $ 0.0 & 1.1 $^{+ 1.3}_{- 0.8}$ & $-2.1 ^{+ 1.1}_{-
  0.1}$ & 0.0 $\pm $ 0.0 & 2.1 $^{+ 0.1}_{- 1.1}$ \\
 
   $\Delta(2390)F_{37}$* & 4.7 $\pm $ 1.0 &
   $<$ 0.9 & & & & & & & \\
 
$[\Delta\textstyle{7\over 2}^+]_3(2460)$ & 1.1 $^{+ 0.0}_{- 0.1}$ & 0.5
  $\pm $ 0.1 & 0.5 $\pm $ 0.4 & 0.8 $\pm $ 0.6 & $-0.1 \pm $ 0.1 & 0.9
  $^{+ 0.7}_{- 0.6}$ & 0.8 $^{+ 0.1}_{- 0.3}$ & $-0.3 \pm $ 0.2 & 0.8
  $^{+ 0.2}_{- 0.3}$ \\
 
& & & $h_{\frac{1}{2}}$ & $f_{\frac{3}{2}}$ & $h_{\frac{3}{2}}$ & &
$f$ & $h$ \\ \cline{4-6} \cline{8-9}

$[\Delta\textstyle{9\over 2}^+]_1(2420)$ & 1.2 $\pm $ 0.4 & 0.2 $\pm $
  0.1 & 0.0 $\pm $ 0.0 & 0.3 $^{+ 0.8}_{- 0.3}$ & 0.0 $\pm $ 0.0 & 0.3
  $^{+ 0.8}_{- 0.3}$ & 0.0 $\pm $ 0.0 & $-0.2 \pm $ 0.2 & 0.2 $\pm $
  0.1 \\
 
   $\Delta(2300)H_{39}$** & 4.8 $\pm $ 1.0 &
   $\simeq$+ 1.4 $\pm $ 0.0 & & & & & & & \\
 
$[\Delta\textstyle{9\over 2}^+]_2(2505)$ & 0.4 $\pm $ 0.1 & 0.1 $\pm $
  0.1 & 0.3 $\pm $ 0.3 & 0.4 $^{+ 0.1}_{- 0.2}$ & $-0.3 \pm $ 0.2 &
  0.6 $\pm $ 0.3 & 1.4 $^{+ 0.6}_{- 0.0}$ & $-0.1 \pm $ 0.0 & 1.4 $^{+
  0.6}_{- 0.0}$ \\
 
& & & $h_{\frac{1}{2}}$ & $h_{\frac{3}{2}}$ & $j_{\frac{3}{2}}$ & &
$h$ & $j$ \\ \cline{4-6} \cline{8-9}

$[\Delta\textstyle{11\over 2}^+]_1(2450)$ & 2.9 $\pm $ 0.7 & 0.5 $\pm $
  0.3 & 0.1 $\pm $ 0.1 & $-0.2 \pm $ 0.1 & 0.0 $\pm $ 0.0 & 0.2 $\pm $
  0.2 & $-0.3 \pm $ 0.1 & 0.0 $\pm $ 0.0 & 0.3 $\pm $ 0.1 \\
 
   $\Delta(2420)H_{3\,\,11}$**** & 6.3 $\pm $
   1.0 & $\simeq$- 1.0 $\pm $ 0.0 & & & & & & & \\
 
& & & $j_{\frac{1}{2}}$ & $h_{\frac{3}{2}}$ & $j_{\frac{3}{2}}$ & &
$h$ & $j$ \\ \cline{4-6} \cline{8-9}

$[\Delta\textstyle{13\over 2}^+]_1(2880)$ & 0.8 $\pm $ 0.2 & 0.1 $\pm $
  0.1 & $-0.1 \pm $ 0.1 & 0.5 $^{+ 0.1}_{- 0.3}$ & $-0.1 \pm $ 0.1 &
  0.5 $^{+ 0.1}_{- 0.3}$ & 0.0 $\pm $ 0.0 & $-0.3 ^{+ 0.2}_{- 0.4}$ &
  0.3 $^{+ 0.4}_{- 0.2}$ \\
 
$[\Delta\textstyle{13\over 2}^+]_2(2955)$ & 0.2 $\pm $ 0.1 & 0.0 $\pm $
  0.0 & 0.3 $\pm $ 0.1 & 0.2 $\pm $ 0.0 & $-0.3 \pm $ 0.2 & 0.4 $\pm $
  0.2 & 1.1 $^{+ 0.3}_{- 0.6}$ & $-0.1 \pm $ 0.1 & 1.1 $^{+ 0.3}_{-
  0.6}$ \\
 
& & & $j_{\frac{1}{2}}$ & $j_{\frac{3}{2}}$ & $l_{\frac{3}{2}}$ & &
$j$ & $l$ \\ \cline{4-6} \cline{8-9}

$[\Delta\textstyle{15\over 2}^+]_1(2920)$ & 1.6 $\pm $ 0.3 & 0.2 $\pm $
  0.1 & 0.2 $\pm $ 0.0 & $-0.3 \pm $ 0.1 & 0.0 $\pm $ 0.0 & 0.3 $\pm $
  0.1 & $-0.4 \pm $ 0.2 & 0.0 $\pm $ 0.0 & 0.4 $\pm $ 0.2 \\
 
   $\Delta(2950)K_{3\,\,15}$** & $3.6\pm 1.5$ \\
 
$[\Delta\textstyle{15\over 2}^+]_2(3085)$ & 0.4 $\pm $ 0.1 & 0.1 $\pm $
  0.0 & 0.0 $\pm $ 0.0 & $-0.1 \pm $ 0.0 & 0.0 $\pm $ 0.0 & 0.1 $\pm $
  0.0 & $-0.2 \pm $ 0.1 & 0.0 $\pm $ 0.0 & 0.2 $\pm $ 0.0 \\
 
\end{tabular}
\end{table}
\begin{table}
\caption{Results in the $\Lambda K$, $\Lambda K^*$, $\Lambda(1405) K$
and $\Lambda(1520) K$ channels for the lightest few negative-parity
$N$ resonances of each $J$ in the N=3 band, and for the lightest few
$N$ resonances for $J^P$ values which first appear in the N=4, 5 and 6
bands. Notation as in Table~\protect{\ref{NDle2SK}}.}
\label{NDge3LK}
\begin{tabular}{lrrrrrrrrr}

 State &  $\Lambda K$ &  $\Lambda K^*$ &  $\Lambda K^*$ & 
  $\Lambda K^*$ &  $\sqrt{\Gamma_{\Lambda K^*}}$ & 
  $\Lambda(1405) K$ &  $\Lambda(1520) K$ &  $\Lambda(1520) K$ & 
  $\sqrt{\Gamma_{\Lambda(1520) K}}$ \\

& & $s_{\frac{1}{2}}$ & $d_{\frac{3}{2}}$ & & & & $p$ & \\ \cline{3-4}
\cline{8-8}

 $[N\textstyle{1\over 2}^-]_3(1945)$ & 2.3 $\pm $ 2.7 & $-2.2 \pm $ 1.7
  & 2.2 $^{+ 4.2}_{- 2.0}$ & & 3.1 $^{+ 4.3}_{- 2.5}$ & 0.5 $^{+
  1.0}_{- 0.4}$ & 6.4 $^{+ 5.7}_{- 6.4}$ & & 6.4 $^{+ 5.7}_{- 6.4}$ \\
 
   $N(2090)S_{11}$*\\
 
 $[N\textstyle{1\over 2}^-]_4(2030)$ & 0.3 $\pm $ 0.5 & $-0.4 ^{+
  0.3}_{- 0.2}$ & 0.1 $^{+ 0.8}_{- 0.1}$ & & 0.4 $^{+ 0.7}_{- 0.4}$ &
  1.2 $^{+ 0.9}_{- 1.1}$ & 0.5 $^{+ 2.2}_{- 0.5}$ & & 0.5 $^{+ 2.2}_{-
  0.5}$ \\
 
 $[N\textstyle{1\over 2}^-]_5(2070)$ & 2.7 $\pm $ 1.3 & $-2.3 ^{+
  2.0}_{- 0.5}$ & 0.9 $^{+ 2.4}_{- 0.8}$ & & 2.4 $^{+ 1.8}_{- 2.1}$ &
  0.1 $\pm $ 0.1 & 1.9 $^{+ 2.3}_{- 1.9}$ & & 1.9 $^{+ 2.3}_{- 1.9}$
  \\
 
 $[N\textstyle{1\over 2}^-]_6(2145)$ & $-0.1 \pm $ 0.1 & 0.0 $\pm $ 0.0
  & $-0.3 ^{+ 0.2}_{- 0.4}$ & & 0.3 $^{+ 0.4}_{- 0.2}$ & 0.0 $\pm $
  0.0 & 1.1 $^{+ 0.6}_{- 1.1}$ & & 1.1 $^{+ 0.6}_{- 1.1}$ \\
 
 $[N\textstyle{1\over 2}^-]_7(2195)$ & $-0.1 \pm $ 0.3 & 0.8 $\pm $ 0.2
  & $-0.8 ^{+ 0.7}_{- 1.1}$ & & 1.1 $^{+ 1.0}_{- 0.5}$ & $-1.0 ^{+
  0.4}_{- 0.0}$ & $-0.7 ^{+ 0.5}_{- 0.1}$ & & 0.7 $^{+ 0.1}_{- 0.5}$
  \\
 
& & $d_{\frac{1}{2}}$ & $s_{\frac{3}{2}}$ & $d_{\frac{3}{2}}$ & & &
$p$ & $f$ \\ \cline{3-5} \cline{8-9}

 $[N\textstyle{3\over 2}^-]_3(1960)$ & $-5.6 ^{+ 1.7}_{- 1.3}$ & 0.7
  $^{+ 1.5}_{- 0.6}$ & 3.8 $\pm $ 2.9 & 1.3 $^{+ 2.9}_{- 1.2}$ & 4.0
  $^{+ 4.1}_{- 3.1}$ & 3.9 $^{+ 1.3}_{- 2.7}$ & $-2.6 ^{+ 2.6}_{-
  2.8}$ & $-0.2 ^{+ 0.2}_{- 1.3}$ & 2.6 $^{+ 2.9}_{- 2.6}$ \\
 
   $N(2080)D_{13}$** & $\simeq + 1.7 \pm $
   1.0 & & & & & & & & \\
 
 $[N\textstyle{3\over 2}^-]_4(2055)$ & $-2.7 ^{+ 0.9}_{- 0.8}$ & 0.2
  $^{+ 0.8}_{- 0.2}$ & 3.3 $^{+ 1.7}_{- 2.9}$ & 0.4 $^{+ 1.6}_{- 0.4}$
  & 3.3 $^{+ 2.1}_{- 3.0}$ & 1.2 $^{+ 0.5}_{- 0.9}$ & $-0.5 ^{+
  0.5}_{- 0.9}$ & 0.0 $^{+ 0.0}_{- 0.5}$ & 0.6 $^{+ 1.0}_{- 0.6}$ \\
 
 $[N\textstyle{3\over 2}^-]_5(2095)$ & $-0.1 \pm $ 0.0 & 0.0 $\pm $ 0.0
  & $-0.6 ^{+ 0.5}_{- 0.2}$ & 0.0 $\pm $ 0.0 & 0.6 $^{+ 0.2}_{- 0.5}$
  & 0.7 $^{+ 0.2}_{- 0.4}$ & 0.4 $\pm $ 0.4 & 0.0 $\pm $ 0.0 & 0.4
  $\pm $ 0.4 \\
 
 $[N\textstyle{3\over 2}^-]_6(2165)$ & 0.2 $\pm $ 0.1 & $-0.1 \pm $ 0.1
  & $-0.1 \pm $ 0.0 & $-0.1 \pm $ 0.1 & 0.2 $\pm $ 0.2 & $-0.1 \pm $
  0.1 & 0.4 $^{+ 0.2}_{- 0.4}$ & 0.0 $\pm $ 0.1 & 0.4 $^{+ 0.2}_{-
  0.4}$ \\
 
 $[N\textstyle{3\over 2}^-]_7(2180)$ & $-0.1 \pm $ 0.0 & 0.1 $\pm $ 0.1
  & $-0.1 \pm $ 0.0 & $-0.1 \pm $ 0.0 & 0.2 $\pm $ 0.2 & 1.5 $^{+
  0.3}_{- 0.7}$ & $-1.1 ^{+ 0.9}_{- 0.4}$ & $-0.1 \pm $ 0.1 & 1.1 $^{+
  0.5}_{- 0.9}$ \\
 
& & $d_{\frac{1}{2}}$ & $d_{\frac{3}{2}}$ & $g_{\frac{3}{2}}$ & & &
$p$ & $f$ \\ \cline{3-5} \cline{8-9}

 $[N\textstyle{5\over 2}^-]_2(2080)$ & $-2.9 ^{+ 0.8}_{- 0.4}$ & 0.9
  $^{+ 1.6}_{- 0.9}$ & $-1.0 ^{+ 0.9}_{- 1.6}$ & $-0.3 ^{+ 0.3}_{-
  1.6}$ & 1.4 $^{+ 2.7}_{- 1.3}$ & 0.1 $^{+ 0.4}_{- 0.1}$ & $-4.7 ^{+
  4.7}_{- 1.2}$ & $-0.3 ^{+ 0.3}_{- 0.8}$ & 4.7 $^{+ 1.3}_{- 4.7}$ \\
 
 $[N\textstyle{5\over 2}^-]_3(2095)$ & $-1.7 ^{+ 0.5}_{- 0.4}$ & 0.2
  $^{+ 0.4}_{- 0.2}$ & $-0.2 ^{+ 0.2}_{- 0.4}$ & 0.0 $^{+ 0.0}_{-
  0.2}$ & 0.3 $^{+ 0.6}_{- 0.3}$ & 0.0 $\pm $ 0.0 & $-2.4 ^{+ 2.4}_{-
  2.0}$ & $-0.1 ^{+ 0.1}_{- 0.3}$ & 2.4 $^{+ 2.0}_{- 2.4}$ \\
 
   $N(2200)D_{15}$** & $\simeq - 2.2 \pm $
   1.0 & & & & & & & & \\
 
 $[N\textstyle{5\over 2}^-]_4(2180)$ & $-0.3 \pm $ 0.1 & 0.1 $\pm $ 0.1
  & $-0.2 \pm $ 0.2 & 0.0 $\pm $ 0.0 & 0.2 $\pm $ 0.3 & $-0.2 \pm $
  0.2 & 0.0 $\pm $ 0.1 & 0.0 $\pm $ 0.0 & 0.0 $\pm $ 0.1 \\
 
 $[N\textstyle{5\over 2}^-]_5(2235)$ & $-0.9 \pm $ 0.2 & 0.3 $^{+
  0.4}_{- 0.2}$ & $-0.3 ^{+ 0.2}_{- 0.5}$ & $-0.1 ^{+ 0.1}_{- 0.4}$ &
  0.4 $^{+ 0.7}_{- 0.3}$ & $-0.1 \pm $ 0.1 & 0.0 $\pm $ 0.1 & $-0.2
  ^{+ 0.2}_{- 0.3}$ & 0.2 $^{+ 0.3}_{- 0.2}$ \\
 
 $[N\textstyle{5\over 2}^-]_6(2260)$ & $-0.2 \pm $ 0.0 & 0.1 $\pm $ 0.1
  & $-0.1 \pm $ 0.1 & 0.0 $\pm $ 0.0 & 0.2 $^{+ 0.2}_{- 0.1}$ & 0.4
  $^{+ 0.5}_{- 0.2}$ & $-2.1 ^{+ 0.9}_{- 0.4}$ & 0.2 $^{+ 0.3}_{-
  0.2}$ & 2.1 $^{+ 0.5}_{- 0.9}$ \\
 
 $[N\textstyle{5\over 2}^-]_7(2295)$ & 0.4 $\pm $ 0.0 & $-0.3 \pm $ 0.1
  & 0.3 $\pm $ 0.2 & 0.1 $^{+ 0.2}_{- 0.1}$ & 0.4 $^{+ 0.4}_{- 0.2}$ &
  0.1 $\pm $ 0.1 & 0.4 $^{+ 0.0}_{- 0.1}$ & $-0.1 \pm $ 0.1 & 0.4 $\pm
  $ 0.0 \\
 
 $[N\textstyle{5\over 2}^-]_8(2305)$ & 0.0 $\pm $ 0.0 &  &
   &  & 0.0 $\pm $ 0.0 & $-0.3 ^{+ 0.2}_{-
  0.4}$ & $-0.3 \pm $ 0.1 & 0.0 $\pm $ 0.0 & 0.3 $\pm $ 0.0 \\
 
& & $g_{\frac{1}{2}}$ & $d_{\frac{3}{2}}$ & $g_{\frac{3}{2}}$ & & &
$f$ & $h$ \\ \cline{3-5} \cline{8-9}

 $[N\textstyle{7\over 2}^-]_1(2090)$ & $-1.3 ^{+ 0.4}_{- 0.6}$ & 0.1
  $\pm $ 0.2 & 2.5 $^{+ 1.0}_{- 1.6}$ & 0.2 $^{+ 0.3}_{- 0.2}$ & 2.5
  $^{+ 1.0}_{- 1.6}$ & 1.2 $\pm $ 0.7 & $-0.5 ^{+ 0.4}_{- 0.6}$ & 0.0
  $\pm $ 0.0 & 0.5 $^{+ 0.6}_{- 0.4}$ \\
 
   $N(2190)G_{17}$**** & $\simeq + 1.1 \pm $ 0.0 & & & & & & & & \\
 
 $[N\textstyle{7\over 2}^-]_2(2205)$ & $-0.5 \pm $ 0.2 & 0.1 $^{+
  0.2}_{- 0.1}$ & 1.0 $^{+ 1.3}_{- 0.8}$ & 0.1 $^{+ 0.3}_{- 0.1}$ &
  1.0 $^{+ 1.3}_{- 0.8}$ & 0.7 $^{+ 0.7}_{- 0.5}$ & $-0.2 ^{+ 0.2}_{-
  0.4}$ & 0.0 $\pm $ 0.0 & 0.2 $^{+ 0.4}_{- 0.2}$ \\
 
 $[N\textstyle{7\over 2}^-]_3(2255)$ & $-0.1 \pm $ 0.1 & 0.0 $\pm $ 0.1
  & 0.3 $^{+ 0.4}_{- 0.2}$ & 0.0 $\pm $ 0.1 & 0.3 $^{+ 0.4}_{- 0.2}$ &
  0.0 $\pm $ 0.0 & $-0.1 \pm $ 0.1 & 0.0 $\pm $ 0.0 & 0.1 $\pm $ 0.1
  \\
 
 $[N\textstyle{7\over 2}^-]_4(2305)$ & 0.1 $\pm $ 0.0 & 0.0 $\pm $ 0.0 &
  $-0.1 \pm $ 0.1 & 0.0 $\pm $ 0.0 & 0.1 $\pm $ 0.1 & $-0.2 ^{+
  0.1}_{- 0.3}$ & 0.2 $\pm $ 0.2 & 0.0 $\pm $ 0.0 & 0.2 $\pm $ 0.2 \\
 
 $[N\textstyle{7\over 2}^-]_5(2355)$ & $-0.3 \pm $ 0.1 & 0.1 $\pm $ 0.0
  & 0.8 $^{+ 0.0}_{- 0.5}$ & 0.1 $\pm $ 0.1 & 0.8 $^{+ 0.0}_{- 0.5}$ &
  0.9 $^{+ 1.0}_{- 0.5}$ & 0.5 $^{+ 0.4}_{- 0.3}$ & 0.0 $\pm $ 0.0 &
  0.5 $^{+ 0.4}_{- 0.3}$ \\
 
& & & & & & & $f$ & $h$ \\ \cline{8-9}

 $[N\textstyle{9\over 2}^-](2215)$ & 0.0 $\pm $ 0.0 & & & & 0.0 $\pm $
   0.0 & 0.0 $\pm $ 0.0 & $-0.2 \pm $ 0.1 & 0.0 $\pm $ 0.0 & 0.2 $\pm
   $ 0.1 \\
 
   $N(2250)G_{19}$**** & $\simeq - 1.2 \pm $
   0.0 & & & & & & & & \\
 
& & $p_{\frac{1}{2}}$ & $p_{\frac{3}{2}}$ & & & &
$d$ & \\ \cline{3-4} \cline{8-8}

 $[N\textstyle{1\over 2}^+]_6(2065)$ & 0.4 $^{+ 1.1}_{- 1.6}$ & $-0.8
  ^{+ 0.7}_{- 0.1}$ & $-2.2 ^{+ 1.9}_{- 0.4}$ & & 2.3 $^{+ 0.4}_{-
  2.0}$ & 5.2 $\pm $ 0.8 & $-1.3 ^{+ 1.3}_{- 2.9}$ & & 1.3 $^{+
  2.9}_{- 1.3}$ \\
 
   $N(2100)P_{11}$* \\
 
 $[N\textstyle{1\over 2}^+]_7(2210)$ & $-0.9 ^{+ 0.3}_{- 0.2}$ & 0.7
  $^{+ 0.3}_{- 0.4}$ & 2.1 $^{+ 0.8}_{- 1.3}$ & & 2.2 $^{+ 0.9}_{-
  1.3}$ & $-0.6 \pm $ 0.3 & 1.0 $\pm $ 0.7 & & 1.0 $\pm $ 0.7 \\
 
& & $f_{\frac{1}{2}}$ & $f_{\frac{3}{2}}$ & $h_{\frac{3}{2}}$ & & &
$d$ & $g$ \\ \cline{3-5} \cline{8-9}

 $[N\textstyle{7\over 2}^+]_2(2390)$ & $-1.7 \pm $ 0.4 & 0.9 $^{+
  0.1}_{- 0.6}$ & $-1.0 ^{+ 0.7}_{- 0.1}$ & $-0.5 \pm $ 0.4 & 1.4 $^{+
  0.2}_{- 1.0}$ & 0.1 $\pm $ 0.0 & 3.1 $^{+ 0.8}_{- 1.2}$ & 0.3 $^{+
  0.3}_{- 0.2}$ & 3.1 $^{+ 0.8}_{- 1.2}$ \\
 
 $[N\textstyle{7\over 2}^+]_3(2410)$ & 0.1 $\pm $ 0.0 & & & & 0.0 $\pm
  $ 0.0 & $-0.1 \pm $ 0.1 & $-0.7 ^{+ 0.2}_{- 0.1}$ & $-0.1 \pm $ 0.0
  & 0.7 $^{+ 0.1}_{- 0.2}$ \\
 
 $[N\textstyle{7\over 2}^+]_4(2455)$ & 0.0 $\pm $ 0.0 & & & & 0.0 $\pm
   $ 0.0 & $-0.2 \pm $ 0.1 & $-0.2 \pm $ 0.1 & 0.0 $\pm $ 0.0 & 0.2
   $\pm $ 0.1 \\
 
& & $h_{\frac{1}{2}}$ & $f_{\frac{3}{2}}$ & $h_{\frac{3}{2}}$ & & &
$g$ & $i$ \\ \cline{3-5} \cline{8-9}

 $[N\textstyle{9\over 2}^+]_1(2345)$ & $-0.4 \pm $ 0.1 & 0.0 $\pm $ 0.1
  & 0.6 $^{+ 0.7}_{- 0.4}$ & 0.0 $\pm $ 0.1 & 0.6 $^{+ 0.7}_{- 0.4}$ &
  $-0.3 ^{+ 0.2}_{- 0.3}$ & 0.1 $\pm $ 0.2 & 0.0 $\pm $ 0.0 & 0.1 $\pm
  $ 0.2 \\
 
   $N(2220)H_{19}$**** & $\simeq$ 0.0 $\pm $
   0.0 & & & & & & & & \\
 
 $[N\textstyle{9\over 2}^+]_2(2500)$ & $-0.2 \pm $ 0.1 & 0.1 $\pm $ 0.1
  & 0.5 $^{+ 0.3}_{- 0.1}$ & 0.1 $^{+ 0.2}_{- 0.0}$ & 0.5 $^{+ 0.3}_{-
  0.1}$ & $-0.6 \pm $ 0.4 & 0.2 $\pm $ 0.1 & 0.0 $\pm $ 0.0 & 0.2 $\pm
  $ 0.2 \\
 
 $[N\textstyle{9\over 2}^+]_3(2490)$ & $-0.1 \pm $ 0.0 & 0.0 $\pm $ 0.0
  & 0.2 $\pm $ 0.1 & 0.0 $\pm $ 0.1 & 0.2 $\pm $ 0.1 & $-0.4 \pm $ 0.2
  & 0.1 $\pm $ 0.1 & 0.0 $\pm $ 0.0 & 0.1 $\pm $ 0.1 \\
 
& & & & & & & $g$ & $i$ \\ \cline{8-9}

 $[N\textstyle{11\over 2}^+]_1(2490)$ & 0.0 $\pm $ 0.0 & & & & 0.0
  $\pm $ 0.0 & 0.0 $\pm $ 0.0 & $-0.2 \pm $ 0.1 & 0.0 $\pm $ 0.0 & 0.2
  $\pm $ 0.2 \\
 
 $[N\textstyle{11\over 2}^+]_2(2600)$ & 0.0 $\pm $ 0.0 & & & & 0.0
  $\pm $ 0.0 & 0.0 $\pm $ 0.0 & 0.4 $\pm $ 0.3 & 0.0 $\pm $ 0.0 & 0.4
  $\pm $ 0.3 \\
 
& & $i_{\frac{1}{2}}$ & $g_{\frac{3}{2}}$ & $i_{\frac{3}{2}}$ & & &
$h$ & $j$ \\ \cline{3-5} \cline{8-9}

 $[N\textstyle{11\over 2}^-]_1(2600)$ & $-0.4 \pm $ 0.2 & 0.1 $^{+
  0.3}_{- 0.1}$ & 0.9 $^{+ 1.2}_{- 0.3}$ & 0.2 $^{+ 0.4}_{- 0.1}$ &
  0.9 $^{+ 1.3}_{- 0.3}$ & 1.0 $\pm $ 0.6 & $-0.4 ^{+ 0.2}_{- 0.4}$ &
  0.0 $\pm $ 0.0 & 0.4 $^{+ 0.4}_{- 0.2}$ \\
 
   $N(2600)I_{1\,\,11}$*** \\
 
 $[N\textstyle{11\over 2}^-]_2(2670)$ & $-0.2 \pm $ 0.1 & 0.1 $\pm $ 0.1
  & 0.4 $^{+ 0.5}_{- 0.2}$ & 0.1 $^{+ 0.2}_{- 0.1}$ & 0.4 $^{+ 0.5}_{-
  0.2}$ & 0.5 $\pm $ 0.3 & $-0.2 \pm $ 0.1 & 0.0 $\pm $ 0.0 & 0.2 $\pm
  $ 0.1 \\
 
 $[N\textstyle{11\over 2}^-]_3(2700)$ & 0.0 $\pm $ 0.0 & 0.0 $\pm $ 0.0
  & 0.1 $\pm $ 0.1 & 0.0 $\pm $ 0.0 & 0.1 $\pm $ 0.1 & 0.0 $\pm $ 0.0
  & 0.0 $\pm $ 0.0 & 0.0 $\pm $ 0.0 & 0.0 $\pm $ 0.0 \\
 
 $[N\textstyle{11\over 2}^-]_4(2770)$ & 0.0 $\pm $ 0.0 & 0.0 $\pm $ 0.0
  & $-0.1 \pm $ 0.1 & 0.0 $\pm $ 0.0 & 0.1 $\pm $ 0.0 & $-0.2 \pm $
  0.1 & 0.1 $\pm $ 0.1 & 0.0 $\pm $ 0.0 & 0.1 $\pm $ 0.1 \\
 
 $[N\textstyle{11\over 2}^-]_5(2855)$ & $-0.1 \pm $ 0.0 & 0.1 $\pm $ 0.0
  & 0.3 $^{+ 0.0}_{- 0.1}$ & 0.1 $\pm $ 0.0 & 0.3 $^{+ 0.0}_{- 0.2}$ &
  0.5 $\pm $ 0.2 & 0.2 $\pm $ 0.1 & 0.0 $\pm $ 0.0 & 0.2 $\pm $ 0.1 \\
 
& & $j_{\frac{1}{2}}$ & $h_{\frac{3}{2}}$ & $j_{\frac{3}{2}}$ & & &
$i$ & $k$ \\ \cline{3-5} \cline{8-9}

 $[N\textstyle{13\over 2}^+]_1(2820)$ & $-0.2 \pm $ 0.1 & 0.1 $\pm $ 0.1
  & 0.5 $^{+ 0.6}_{- 0.3}$ & 0.1 $^{+ 0.2}_{- 0.1}$ & 0.5 $^{+ 0.6}_{-
  0.3}$ & $-0.5 \pm $ 0.2 & 0.2 $\pm $ 0.2 & 0.0 $\pm $ 0.0 & 0.2 $\pm
  $ 0.2 \\
 
   $N(2700)K_{1\,\,13}$** \\
 
 $[N\textstyle{13\over 2}^+]_2(2930)$ & $-0.1 \pm $ 0.0 & 0.1 $\pm $ 0.0
  & 0.3 $\pm $ 0.0 & 0.1 $\pm $ 0.0 & 0.3 $\pm $ 0.0 & $-0.3 \pm $ 0.1
  & 0.1 $\pm $ 0.1 & 0.0 $\pm $ 0.0 & 0.1 $\pm $ 0.1 \\
 
 $[N\textstyle{13\over 2}^+]_3(2955)$ & 0.0 $\pm $ 0.0 & & & & 0.0
  $\pm $ 0.0 & $-0.1 \pm $ 0.0 & & & 0.0 $\pm $ 0.0 \\
 
& & & & & & & $h$ & $j$ \\ \cline{8-9}

 $[N\textstyle{13\over 2}^-]_1(2715)$ & 0.0 $\pm $ 0.0 & & & & 0.0
  $\pm $ 0.0 & 0.0 $\pm $ 0.0 & $-0.1 \pm $ 0.0 & 0.0 $\pm $ 0.0 & 0.1
  $\pm $ 0.1 \\
 
 $[N\textstyle{13\over 2}^-]_2(2845)$ & 0.0 $\pm $ 0.0 & & & & 0.0
  $\pm $ 0.0 & 0.0 $\pm $ 0.0 & 0.2 $\pm $ 0.1 & 0.0 $\pm $ 0.0 & 0.2
  $\pm $ 0.1 \\
 
& & & & & & & $i$ & $k$ \\ \cline{8-9}

 $[N\textstyle{15\over 2}^+]_1(2940)$ & 0.0 $\pm $ 0.0 & & & & 0.0
  $\pm $ 0.0 & 0.0 $\pm $ 0.0 & $-0.1 \pm $ 0.0 & 0.0 $\pm $ 0.0 & 0.1
  $\pm $ 0.0 \\
 
 $[N\textstyle{15\over 2}^+]_2(3005)$ & 0.0 $\pm $ 0.0 & & & & 0.0
  $\pm $ 0.0 & 0.0 $\pm $ 0.0 & 0.2 $\pm $ 0.1 & 0.0 $\pm $ 0.0 & 0.2
  $\pm $ 0.1 \\

\end{tabular}
\end{table}

\end{document}